\documentclass[12pt,preprint]{aastex}
\slugcomment{Version: \today}

\shorttitle{NIR Spectra of Stars in Cyg~OB7}
\shortauthors{Aspin \& }

\begin{document}

\title{NEAR-IR SPECTROSCOPY OF YOUNG STARS IN THE BRAID NEBULA STAR FORMATION REGION IN CYGNUS~OB7}

\author{Colin~Aspin}
\affil{Institute for Astronomy, University of Hawaii, 640 N. A'ohoku Place, Hilo, HI 96720 \\
  {\it caa@ifa.hawaii.edu}}

\author{Tracy~L.~Beck}
\affil{Space Telescope Science Institute, 3700 San Martin Drive, Baltimore, MD 21218 \\
  {\it tbeck@stsci.edu}}

\author{Tae-Soo~Pyo}
\affil{Subaru Telescope, National Astronomical Observatory of Japan, 650 N. A'ohoku Place, Hilo, HI 96720 \\
  {\it pyo@subaru.noaj.org}}

\author{Chris J.~Davis, Gerald M.~Schieven}
\affil{Joint Astronomy Centre, 660 N. A'ohoku Place, University Park, Hilo, Hawaii 96720 \\
  {\it c.davis@jach.hawaii.edu, g.schieven@jach.hawaii.edu}}

\author{Tigran~Khanzadyan}
\affil{Centre for Astronomy, School of Physics, National University of Ireland 
Galway, Galway, Ireland \\
  {\it tigran.khanzadyan@nuigalway.ie}}

\author{Tigran~Yu.~Magakian, Tigran~A.~Movsessian, Elena~G.~Nikogossian}
\affil{V. A. Ambartsumyan Byurakan Astrophysical Observatory, Armenia \\
  {\it tigmag@sci.am. tigmov@bao.sci.am, elena@bao.sci.am}}

\author{Sharon~Mitchison, Michael~D.~Smith}
\affil{Centre for Astrophysics \& Planetary Science, School of Physical Sciences, The University of Kent, Canterbury CT2 7NR, England \\
  {\it smm23@kent.ac.uk, m.d.smith@kent.ac.uk}}

\begin{abstract} 
We present 1.4 to 2.5~$\mu$m integral field spectroscopy of 16 stars in the Braid Nebula star formation region in Cygnus~OB7.  These data forms one aspect of a large-scale multi-wavelength survey aimed at determining an unbiased estimate of the number, mass distribution, and evolutionary state of the young stars within this one square degree area of the previously poorly studied Lynds~1003 molecular cloud.  Our new spectroscopic data, when combined with 2MASS near-IR photometry, provide evidence of membership of many of these objects in the regions pre-main sequence population.  We discuss both the characteristics of the young stars found in the region and the level of star forming activity present. 
\end{abstract}


\section{INTRODUCTION}

The Cygnus~OB7 association is part of the extensive Cygnus molecular cloud complex which contain the more commonly studied Cygnus~X, North American, and Pelican Nebulae regions.   The recent comprehensive review of Reipurth \& Schneider (2008) gives extensive background information on the whole area.   Cygnus~OB7 is one of the nine OB associations in Cygnus and is generally considered the nearest at a distance of around 800~pc (Hiltner 1956; Schmidt 1958; de Zeeuw et al. 1999).  Cygnus~OB7 contains numerous dark cloud complexes, identified with Lynds catalog numbers (Lynds 1962), and forms part of the large-scale dark cloud designated Kh~141 (Khavtassi 1960).  One Lynds cloud, L~1003, was first investigated in detail by Cohen (1980) who found a red nebulous object (designated RNO~127) which was later identified as a bright Herbig-Haro (HH) object (HH448, Melikian \& Karapetian 2001, 2003).  The region was the focus of optical/near-IR (NIR) studies by Devine, Reipurth, \& Bally (1997) and Movsessian et al. (2003).   Both groups found a number of additional HH objects in the region and, together with the presence of numerous IRAS sources, confirmed that L~1003 was the site of significant star formation activity.  In addition, Movsessian et al. (2006) confirmed the presence of a new FU~Orionis eruptive variable (FUor, Herbig 1989) in the cloud.  This object is not optically visible and the outburst was observed in the NIR by the appearance of a bright point-source and extensive reflection nebulosity between the 2MASS survey (taken in 1999) and the acquisition of the Movsessian et al. (2003) images (taken in 2001).  They interpreted the morphology of the NIR nebula as being a double-spiral or ``braid'', created in the walls of both high- and low-velocity outflow cavities, and the object has henceforth been referred to as the Braid Nebula.  The illuminating star of the Braid Nebula will henceforth be referred to as the Braid star since an identification in the General Catalog of Variable Stars (GCVS), is currently pending (Samus, GCVS Newsletter~80, private communication).  

The relatively few previous observations of L~1003 have served to show that it is worthy of a more intensive study.  To this end, we have undertaken an campaign to investigate the young stellar population of the region over a wide-range of wavelengths and using a variety of techniques.   The multi-discipline consortium of researchers involved in this work (the authors on this paper) allow us to cover observational studies from x-rays to the mm/sub-mm.  Our region of focus is a one square degree area of L~1003 centered on the Braid star.  This region actually contains two FUor-like objects since the object HH381~IRS was also found to share most of the characteristics of the small FUor population (Reipurth \& Aspin 1997; Magakian et al. 2008).   The focus of other papers will be optical and NIR broad-band photometry of the stars in the field and extinction through the cloud (Smith et al. 2008; Mitchison et al. 2008), the optical and NIR shock excited line emission features present (Magakian et al. 2008; Khanzadyan et al. 2008), a focused study of the NIR [\ion{Fe}{2}] emission from HH381~IRS (Pyo et al. 2008), and sub-mm CO 3-2 and HCN molecular line maps of the Braid star and HH381~IRS regions (Schieven et al. 2008).   In this paper, we concentrate on NIR spectroscopic observations of a number of relatively bright objects in the study region with the aim of characterizing their evolutionary state, physical properties, and accretion activity.

A NIR spectroscopic survey is a powerful technique to use when attempting to determine the above characteristics of young stellar objects (YSOs).  Such surveys have previous been made in a number of star formation regions, for example, $\rho$~Oph (e.g. Greene \& Lada 1996), NGC~1333 (e.g. Aspin 2003; Wilking et al. 2004), and Cha~I (G\`omez \& Mardones 2003) producing results that allow the estimation of luminosity, mass, age, and accretion properties.  In $\S 2$ below, we present our new observations while in $\S 3$ we discuss what is known about each source observed, together with details of their photometric and spectroscopic properties.  In $\S 4$ we consider the nature of these young stars in relation to the global properties of our survey region.

\section{Observations and Data Reduction}

Spectroscopic observations of 16 stars in our one degree survey region of L~1003 are presented below.  They were obtained on UT 2007 July 07 and 08 on the United Kingdom InfraRed Telescope (UKIRT) using the facility instrument UIST (Ramsey-Howat et al. 2004).  UIST is a 1--5~$\mu$m imager-spectrometer utilizing a 1024$\times$1024 ALADDIN InSb detector array at a pixel scale of 0$\farcs$12~pixel$^{-1}$.  We used the `HK' grism and the Integral Field Unit (IFU) to obtain spectroscopy from 1.4 to 2.5~$\mu$m.  The IFU utilizes an image slicing design with a field-of-view of 3$\farcs$3$\times$6$''$ resulting in a micro-mirror defined spatial resolution of 0$\farcs$12$\times$0$\farcs$24~pixel$^{-1}$ and, with the HK grism, a spectral resolution of R$\sim$900.  Our total exposure time ranged from 60 to 300 seconds for a source brightness of 9$\le$m$_K\le$12. Separate sky observations were obtained adjacent to the target observations (in an ABA pattern where A is target and B is sky) for the accurate subtraction of sky emission lines.  Wavelength calibration was performed using observations of the internal Ar lamp using the same instrument setup as for the on-sky observations.

Basic data reduction was performed automatically in the UKIRT ORACDR pipeline (Cavanagh et al. 2008) using the reduction recipe {\tt MAP\_EXTENDED\_SOURCE}. This resulted in reduced data cubes for each source, sky-subtracted and ready for telluric correction.  Point-source spectra were extracted from the target and telluric standard data cubes using the FIGARO (Shortridge et al. 2002) program {\tt xtplane}.  Point-source 1-dimensional spectra were optically extracted from the resulting 2-dimensional images (equivalent to a long-slit spectroscopic image) using the FIGARO programs {\tt profile} and {\tt optextract}. Telluric correction was performed using the SpeXTool (Cushing, Vacca, \& Rayner 2004) IDL program {\tt xtellcor\_general}.  The telluric standards used were all of spectral type A0~V and observed close in airmass and time to the target observations.  The intrinsic Br$\gamma$ absorption line in the standards was removed via linear interpolation prior to use.

\section{RESULTS}

\begin{figure*}[tb]
\epsscale{1.1}
\plotone{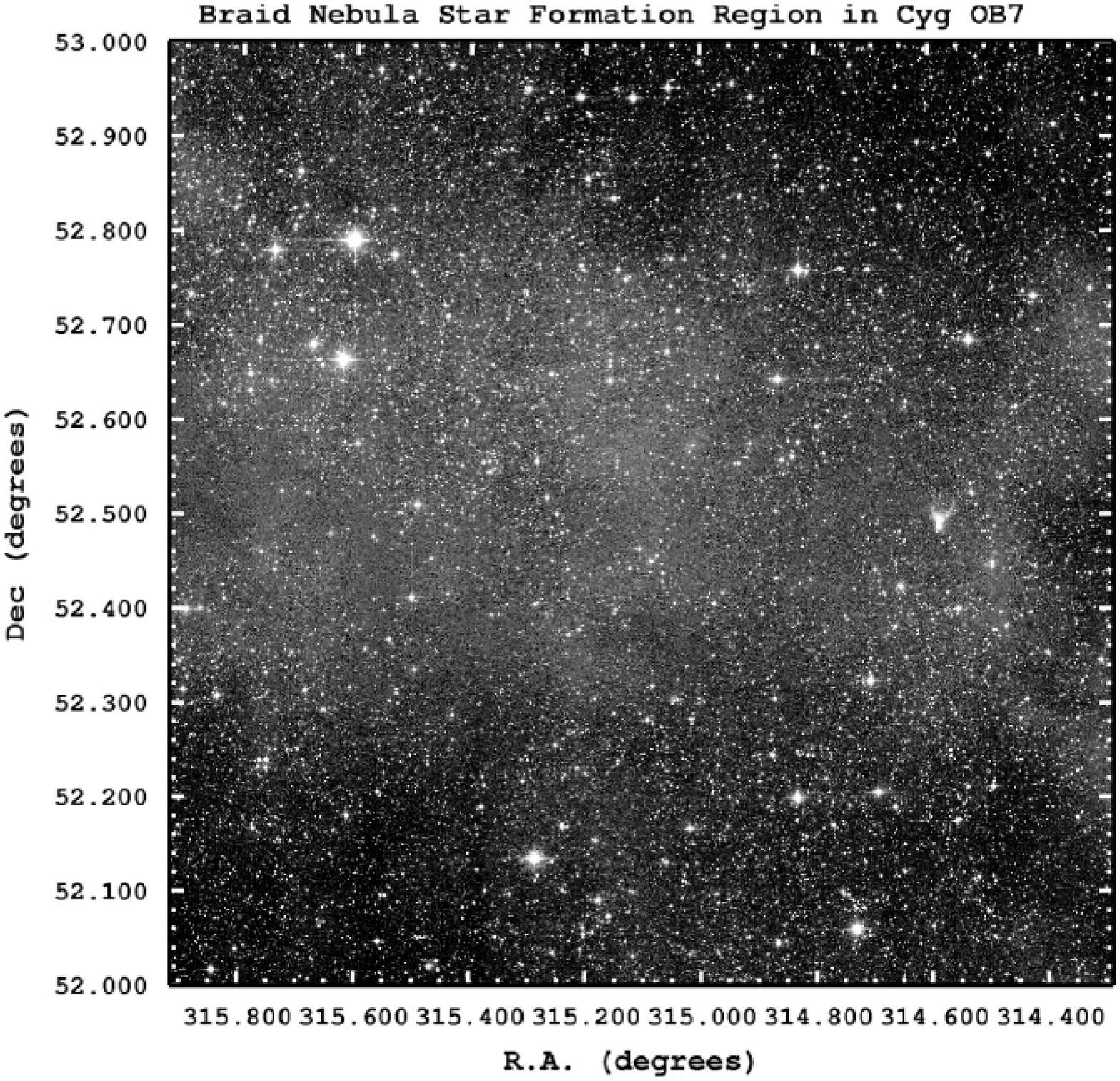} 
\caption{A color composite image of the one square degree Braid Nebula star formation region in Cyg~OB7.  The 2MASS J, H, and K image of the region have been combined as blue, green, and red, respectively.  The sources we have observed spectroscopically are distributed throughout the molecular cloud material spanning the image from east to west.  HH381~IRS can be seen at approximately 314$\farcs$6,52$\farcs$5.  The Braid Nebula itself is located at the center of this image but at the time the 2MASS images were taken it was not visible even at 2~$\mu$m.
\label{region}}
\end{figure*}

Our earlier UKIRT/WFCAM JHK imaging of the approximately 1$^{\circ}$ region of Cygnus~OB7 centered on the Braid Nebula together with the earlier near-IR (NIR) imaging presented in Movsessian et al. (2003) showed the presence of a number of bright K-band sources.  In addition, K-band detections of several IRAS sources were made.  The sources observed with UIST, together with 2MASS identifications and coordinates, are listed in Table~\ref{sourceci}.  Table~\ref{sourceph} gives the J, H, and K photometry of these sources from both 2MASS and our (2006) WFCAM images.  We have not transformed the WFCAM JHK photometry onto the 2MASS photometric system\footnotemark\footnotetext{See http://casu.ast.cam.ac.uk/documents/wfcam/photometry/index for details} and so  difference of up to a tenth of a magnitudes can be expected.  Fig.~\ref{region} shows a composite 2MASS JHK image of the whole one square degree region of study.  Our source sample has K band apparent magnitudes, m$_K$, ranging from 8.3 to 14.1 and NIR J-K colors from 0.4 to 5.9.  Due to the relatively bright nature of many of the sources, however, some of our WFCAM observations were saturated.  In Fig.~\ref{ids1} we identify the target objects by showing the regions surrounding them at optical (Subaru Superime-Cam R$_C$ images) and NIR (UKIRT WFCAM K images) wavelengths.  Our WFCAM images have 5$\sigma$ limiting magnitudes of approximately, 19, 18.5, and 18 at J, H, and K, respectively.  All but two of the sources observed spectroscopically are optically visible. These two sources are the Braid star and IRAS~20591+5214.  Fig.~\ref{ids1} shows that three of the compact nebula (CN) sources (CN~1, CN~3, CN~6) in this field have relatively bright nearby K band stars.  In these cases, we have defined the second stars (not associated with the compact nebula) with either the suffix N (northern) or S (southern), since in all three they are aligned approximately north--south.  In total therefore, with these second stars, we have spectroscopically observed 16 objects in this region.  The J-H vs. H-K color--color diagram (henceforth JHKcc) for all sources with 2MASS photometry is shown in Fig.~\ref{ccdiag}.  

In Fig.~\ref{sp-1} we present the H and K band spectra of the 16 stars observed.  However, since we used an IFU for these observations and hence had spatially resolved spectroscopy, we first studied the immediate environment of all stars to look for possible close companions and extended nebulosity such as HH jets and reflection nebulae.  Only in the cases of CN~1, CN~8 and HH381~IRS did we find anything other than a point-source.  CN~1 has faint [\ion{Fe}{2}] emission extending several arcseconds to the north and possibly to the south of the star. CN~8 shows a nebulous knot at $+\sim$1$''$,$+\sim$1$''$ with respect to the star. This appears to be a continuum source and could be either a fainter companion star or a small reflection knot.  HH381-IRS shows evidence for extended [\ion{Fe}{2}] emission at 1.644~$\mu$m which takes the form of an approximately north-south jet that can be traced to 2$''$ north and 1$''$ south of the star.  We defer more discussion of the [\ion{Fe}{2}] extended emission to a future paper (Pyo et al. 2009) where we will consider them in greater detail. All other IFU datacubes showed point-sources only and hence, we optimally combined the signal from all IFU elements containing flux to obtain the spectra shown in Fig.~\ref{sp-1}.  The spectra have been normalized to unity at 2.2~$\mu$m and the major atomic and molecular features are indicated.  Figs.~\ref{sp-1-h} and \ref{sp-1-k} show closeups of the K and H band region of the spectra, respectively.  We show these plot so that the lines/bands present can be seen in closer detail and also so that the weaker spectral features can be identified seen on the generally strong continuum. Table~\ref{whatp} details the spectral features present in the target sources.

The morphological and absorption/emission line structure of the 16 stars observed spectroscopically are described below.  We refer to their NIR photometry and colors which are shown in Table~\ref{sourceph} and Fig.~\ref{ccdiag} and also their spectral features in the H and K band.  The lines and bands are listed in Table~\ref{whatp} together with their equivalent widths, W$_{\lambda}$.  Also given in this Table are the signal to noise values achieved in the final extracted one-dimensional point-source spectra.  In the discussion below, we repeatedly refer to several atomic absorption lines and molecular bands by their element and transition, e.g. \ion{Ca}{1}.  These are Br$\gamma$ at 2.166~$\mu$m, \ion{Na}{1} at 2.206~$\mu$m, \ion{Ca}{1} at 2.264~$\mu$m, \ion{Mg}{1} at 2.281~$\mu$m, and the CO overtone bandheads longward of 2.294~$\mu$m.  The wavelength of any other features discussed will be defined explicitly.  In the spectra presented above, some of these features are found to appear in emission rather than absorption.   In low-mass stars ($<$2~M$_{\odot}$), Br$\gamma$ emission is typically associated with emission from an accretion flow.  In more massive stars (2$<$M$_{\odot}<$8), it can also be from an associated \ion{H}{2} region.  In higher mass stars, \ion{H}{2} emission can be additionally associated with stellar winds.  CO bandhead emission is thought to occur in dense gas in the inner regions of the circumstellar disk.  We also comment on the state of water vapor absorption in these spectra.   The dominant water band in the HK spectra spans the wavelength range 1.7 to 2.2~$\mu$m and is identified and best seen in the spectral plots (Fig.~\ref{sp-1}).   Finally, the acronyms CTTS and ZAMS refer to classical T~Tauri star and zero-age main sequence, respectively.

In addition to a qualitative description of the spectra, we have also produced best-fits of template star spectra to some of the stars observed i.e. those with good signal to noise (S:N$>$20 in both H and K) and which show strong enough atomic and molecular absorption features.   The modeling procedure was described in detail in Aspin, Beck, \& Reipurth (2008) and based on the Prato, Greene, \& Simon (2003) model extended by Beck (2007).  Briefly, a $\chi^2$ minimization search is performed over the the parameters, stellar spectral type, K-band veiling (r$_{K}$), visual extinction (A$_v$), effective dust temperature (T$_{dust}$), and ice band optical depth ($\tau_{ice}$) and results in estimates of the stellar spectral type, effective temperature of the dust component (T$_{dust}$), visual extinction (A$_V$), and K-band veiling (r$_K$).  The spectral templates are taken from the NASA IRTF SpeX stellar library.\footnotemark\footnotetext{http://irtfweb.ifa.hawaii.edu/$\sim$spex/WebLibrary/} Values for the above physical parameters are given below in Table~\ref{whatn}.

\subsection{The Braid Star}
The Braid Nebula is located at the center of our 1$^{\circ}$ square survey region.  It is so-named due to the optical morphological structure in the images of Movsessian et al. (2006), which resembled a double-helix or ``braid''.  In that paper, Movsessian et al. presented considerable evidence for the young star illuminating the Braid Nebula being of an eruptive variable of the FU~Orionis (FUor) type.  In particular, i) its associated optical nebulosity had brightened considerably between the DSS2 sky survey plate of the region (1990) and their optical images (2001), ii) it had brightened in the NIR by 5 magnitudes between the 2MASS K-band image (1999) of the region and their K-band images (2001), iii) it showed optical absorption features similar to those of FU~Ori itself, e.g. a P~Cygni H$\alpha$ profile and strong \ion{Ba}{2} absorption (at 6498~\AA), and iv) it possessed deep NIR CO overtone bandhead absorption seen also in FU~Ori and characteristic of a cool, low surface gravity atmosphere.  At K, the Braid star exhibited a V-shaped structure centered on the star (typical of scattering in outflow cavity walls) and a long curving comma-like tail (see Fig.~\ref{ids1}).  Strangely however, it was found not to be coincident with a mid-IR (MIR) MSX, nor far-IR (FIR) IRAS, point-source.  

The Braid Star only brightened to a detectable level in the NIR after the 2MASS survey data had been acquired and hence does not have a 2MASS designation nor 2MASS photometry.  Post-2MASS, it was so bright as to be saturated in our WFCAM images (having a saturation limit of m$_K\sim$11.5).  The only K-band  photometry of the source post-outburst is that given in Movsessian et al. (2006) which states m$_K$=10.45$\pm$0.5.  Using the WFCAM J and H band values and the K-band saturation limit, we can derive NIR colors and investigate the location of the Braid star on a NIR color-color diagram.  With a J-H$\sim$1.1 and H-K$>$3.8, it is located beyond the limits of our color-color diagram shown in Fig.~\ref{ccdiag}, however, its clearly shows that the Braid star possesses a large K-band thermal excess. 

A 1.3 to 4.0~$\mu$m spectrum of the Braid star was published in Movsessian et al. (2006).  It showed a very red continuum with strong CO overtone absorption, water vapor absorption from 1.7 to 2.2~$\mu$m, deep water ice absorption at 3~$\mu$m and little else. There was no evidence for strong photospheric \ion{Na}{1}, \ion{Ca}{1}, and \ion{Mg}{1} absorption features leading Movsessian et al. to conclude that the Braid star appeared like an FUor in the NIR.  Our current H and K-band spectrum was taken around 2 years after that of Movsessian et al. and shows very similar features, again, with significant water vapor absorption.  The expanded views of the H and K-bands shown in Figs.~\ref{sp-1-h} and \ref{sp-1-k} perhaps shows very weak \ion{Na}{1}, \ion{Ca}{1}, and \ion{Mg}{1} absorption but certainly shows strong CO overtone absorption (see Table~\ref{whatp} for the line/band W$_{\lambda}$ values).  The presence of water vapor absorption suggest a relatively cool atmosphere since these bands are only found in late-type M dwarfs and giants with effective temperatures, T$_{eff}\lesssim$3500~K.  

The Braid star is certainly a young object and it appears be in an FUor eruptive state.  In the NIR, it has a large thermal excess and spectroscopically (due to its deep CO absorption) resembles an M-type giant/supergiant.  It also drive shock excited Herbig-Haro (HH) flows seen in both the optical (Magakian et al. 2008) and NIR (Khanzadyan et al. 2008), and has considerable associated reflection nebulosity (see Fig.~\ref{ids1}).  It is not detected in the MIR (by MSX) nor in the FIR (by IRAS, except at 25~$\mu$m) suggesting the dust content of the circumstellar region is relatively small, however, it is highly extinguished in the optical implying it is perhaps viewed through a large optical depth of cold obscuring material.  In addition, it does not appear to drive an active molecular CO outflow nor possesses a significant molecular gas reservoir in its immediate vicinity (see the CO and HCN maps in Schieven et al. 2008).  The Braid star should probably be classed as an embedded YSO with an age, $t$, around a few $\times$10$^6$~years, although it is somewhat troublesome to reconcile its optical obscured nature and large K-band thermal excess, with no MIR nor FIR emission and no molecular gas detection.  One interpretation is that perhaps the source has evolved enough to have exhausted its circumstellar dust/gas envelope and much of its cooler circumstellar disk except for the hotter inner regions (where the K-band thermal excess originates).  However, the question arises as to how the disk material has been removed while the star is still seemingly in a young evolutionary state.   No Br$\gamma$ emission is observed in the NIR which may be explained by the disk dominating the NIR spectrum and it is the disks cool, low-surface gravity atmosphere that is responsible for the deep CO overtone absorption.   Detections of the Braid star in the thermal-IR and at sub-mm/mm wavelengths may help resolve these issues.

\subsection{CN~1}
The compact nebulous object designated CN~1 by Movsessian et al. (2003) is the optical counterpart of IRAS~20590+5221.  In the optical, it exhibits a monopolar cavity-like morphology (likely seen in reflected light) which extends $\sim$40$''$ in a northerly direction from the star (see Fig.~\ref{ids1}) with the western wall appearing significantly brighter than its eastern counterpart.  Between the curving cavity walls, a small-scale ($<$10$''$) optical jet is present and has been designated HH632 (Magakian et al. 2008).  About 30$''$ south of CN~1 is a source of similar optical brightness but considerably brighter in the NIR ($>$2 magnitudes brighter at K).  This star shows no evidence of extended emission structure.  We henceforth refer to this object as CN~1S and is considered separately below.

CN~1 is moderately reddened  with a J-K$\sim$2.7 and in the JHKcc diagram (Fig.~\ref{ccdiag}) it has a clear NIR thermal excess.  We also find that CN~1 appears to have remained at about the same NIR brightness over the period 1999 (2MASS photometry) to 2006 (WFCAM photometry). 

The NIR H and K spectrum of CN~1 (Fig.\ref{sp-1}) shows broad water vapor absorption  most noticeably from 1.7 to 2.2~$\mu$m.  It also possesses weak Br(10-4) emission at 1.737~$\mu$m and Br$\gamma$ emission.  The H emission confirms this object as a young actively accreting star while the water vapor absorption suggests the presence of a cool atmosphere.  Weak \ion{Ca}{1} and perhaps very weak \ion{Na}{1} absorption are present as are weak CO overtone absorption bands (best seen in Fig.~\ref{sp-1-k}).  The strength of these features are however insufficient for template model fitting.  [\ion{Fe}{2}] (1.644~$\mu$m) emission is detected (probably from the HH jet) but no obvious H$_2$ emission is seen.  
  
The fact that CN~1 possesses a significant thermal excess, is optically visible, has both a outflow cavity and compact HH jet, and is associated with an (albeit weak) IRAS source, confirms that it is a young star with a circumstellar disk with a hot inner region (responsible for the large K-band thermal excess).  The weak Br$\gamma$ emission suggests that it is actively, but not strongly, accreting.  If HH ejection is tied to accretion rate (in say, an outburst event), the presence of a compact and collimated HH jet would suggest that it has undergone a period of more significant accretion in the recent past.  Assuming a typical HH jet velocity of $\sim$200~km~s$^{-1}$ and an inclination with respect to the plane of the sky of 45$^\circ$, the jet would have been initiated around 300 years ago.  The presence of the monopolar cavity structure suggests that, at some time in the past, CN~1 had an active wide-angle molecular outflow that may still be active.   We classify CN~1 as an accreting CTTS star.

\subsection{CN~1S}
CN~1S has a J-K$\sim$1.3 and in the JHKcc diagram (Fig.~\ref{ccdiag}) its location is consistent with it being either a background (A$_V\sim$6) early-type main sequence dwarf or a reddened CTTS.  

The H and K band spectrum of CN~1S shows a declining (to the red) continuum with strong H absorption lines from Br$\gamma$ down to the Br(15-4) at 1.57~$\mu$m.  In addition, there is no evidence for \ion{He}{1} absorption (1.7~$\mu$m) nor water vapor absorption.  The presence of H absorption features and absence of He features on an otherwise featureless continuum plus its location in the color-color diagram implies that CN~1S is a reddened early-type main sequence star and not a young star.  The expanded K band spectrum of CN~1S in Fig.~\ref{sp-1-k} is a typical early-type spectrum (similar to the telluric calibrators we have used).  With no \ion{He}{1} absorption, CN~1S has to be a B8~V star or later.  Assuming CN~1S is an B8~V star with M$_K$=0.1 magnitudes and reddened by A$_V$=6 (A$_K$=0.6) this would place the star at a distance of $\sim$800pc.  This is the minimum distance  consistent with it being a background object since a later spectral type would result in CN~1S being foreground to the L1003 molecular cloud.  

\subsection{CN~2}
CN~2 was also found in the Movsessian et al. (2003) investigation of the region around RNO~127.   It appeared, as the name implies, compact and nebulous in their optical images with distinctly fan-like morphology.  It is associated with a collimated HH flow, HH630, which extends $\sim$35$''$ to the south-east.  Our new optical images from Subaru, however, show this is actually a considerably longer collimated flow stretching $\sim$72$''$ to the south-east.  In addition, a second HH flow is seen emanating from CN~2.  This is also collimated, extends to the north-west by $\sim$67$''$, and terminates in HH~670. The two flows are not co-linear with the south-east flow being angled more to the north with an angular offset between the south-east and north-west flows of $\sim$10$^{\circ}$.  The north-east flow is more pronounced in H$\alpha$ than [\ion{S}{2}], in fact, in [\ion{S}{2}] only HH670 is seen.  The south-east flow is visible in both H$\alpha$ and [\ion{S}{2}].  The reader is referred to Magakian et al. (2008) for more information on these flows.  CN~2 itself appears to be a single star in our Subaru images suggesting that if it is a binary, with both components driving HH flows, they must be quite close.  In the near-IR, CN~2 also appears single.

The location of CN~2 in the JHKcc diagram (Fig.~\ref{ccdiag}) indicates that it possesses a K-band thermal excess and it appears close to the Meyer, Calvet, \& Hillenbrand (1997, henceforth MCH97) CTTS locus.  The source was saturated in our WFCAM H and K images but not in J. The WFCAM (2006) and 2MASS (1999) J-band photometry are different by only 0.15 magnitude suggesting that, within the photometric uncertainties and transformation differences, the star is not obviously variable.  However, optical photometry from 2000 (Magakian, unpublished data) has shown CN~2 to be variable by up to one magnitude.

\begin{figure*}[tb] 
\epsscale{1.0}
\plotone{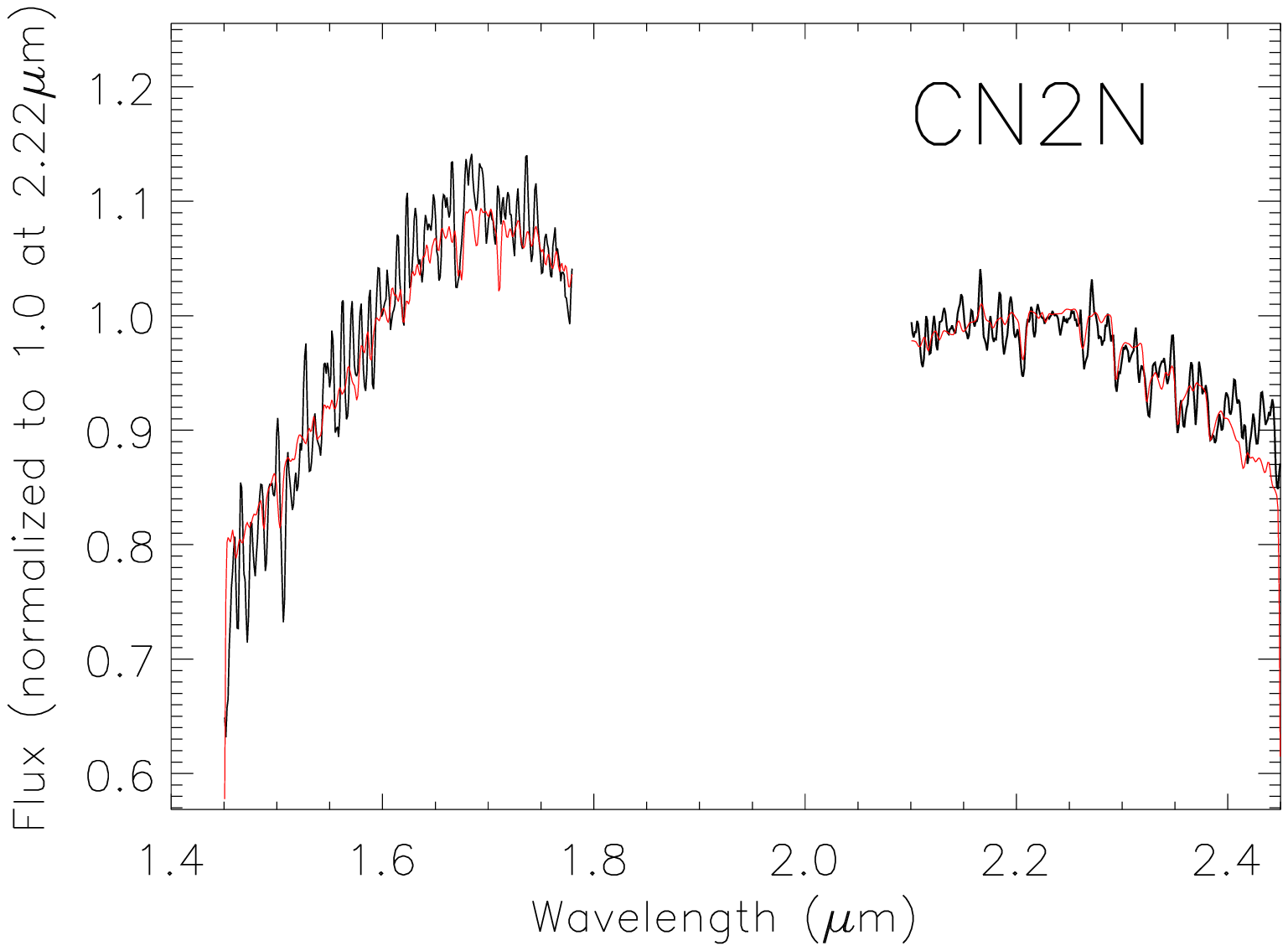} 
\caption{Plot of the observed CN~2 spectrum (in black) overlaid with the best-fit spectral template determined by $\chi^2$ minimization (in red).  The free parameters of the fit are spectral type, dust temperature (T$_{dust}$), visual extinction (A$_V$), and K-band veiling (r$_K$).   The best-fit was obtained for a spectral type M2$\pm$2, A$_V$=12$\pm$2, and r$_K$=0.4$\pm$0.2.\label{bestfit}}
\end{figure*}

\begin{figure*}[tb] 
\epsscale{1.0}
\plotone{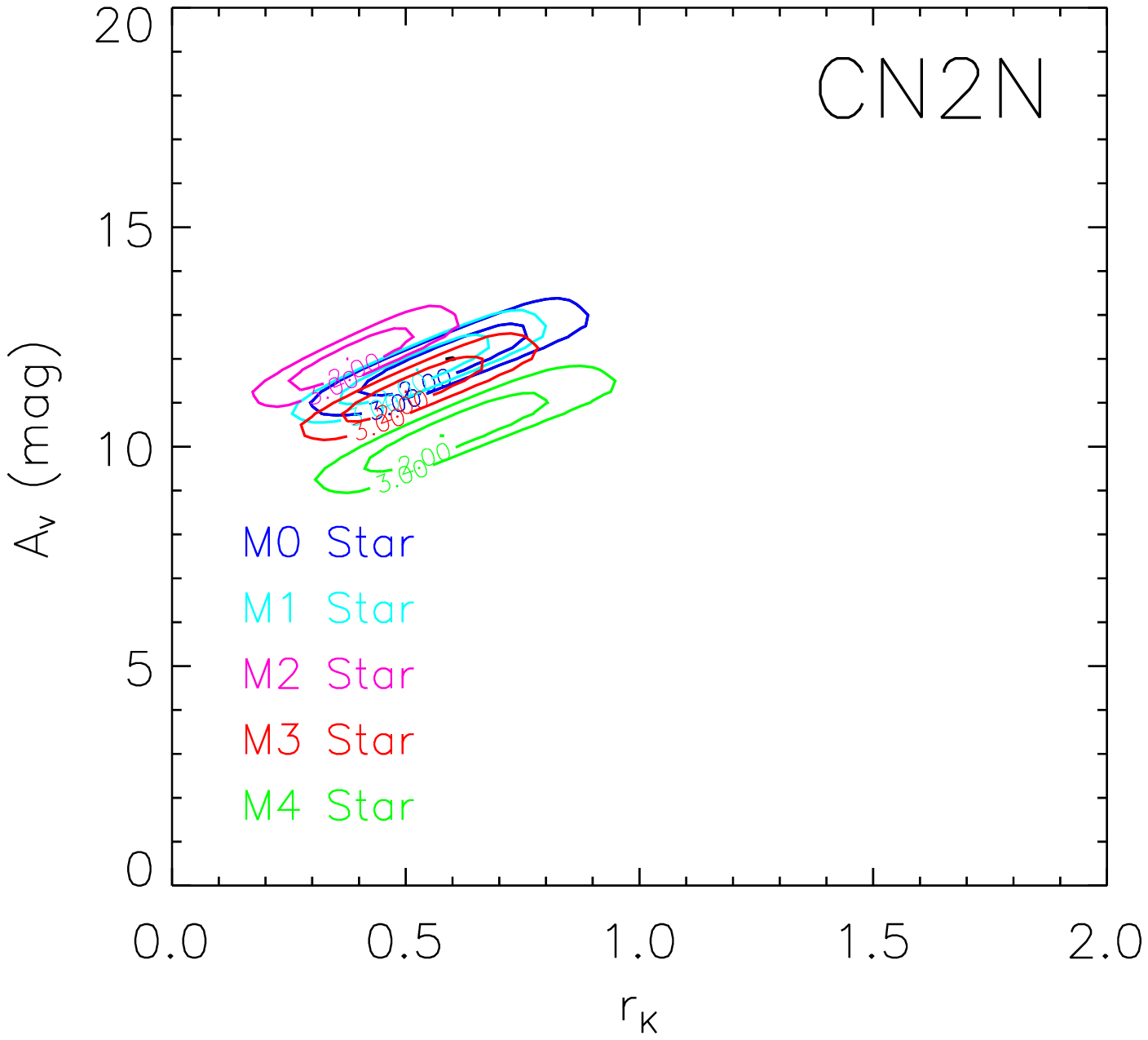}\vspace{1cm}
\caption{Plot of the $\chi^2$ surfaces obtained from the minimization procedure.  This plot from the best-fitting of the spectral template models to the CN~2 spectrum.  Surfaces of 1.01, 2.0, and 3.0 are shown for several spectral types in the A$_V$ vs. r$_K$ plane.  $\chi^2$=3.0 corresponds to approximately 1$\sigma$.  Blue surfaces are for spectral type M0, cyan surfaces for M1, magenta surfaces for M2, red surfaces for M3, and green surfaces for M4.
\label{chisq}}
\end{figure*}

The H and K band spectrum of CN~2 shows a turn-over at $\sim$1.7~$\mu$m.  Such a feature can be attributed to the effects of significant visual extinction.   Weak water vapor absorption is likely present (indicated by the dip in the continuum level longward of 2.3~$\mu$m) together with several weak absorption features, specifically \ion{Na}{1} and \ion{Ca}{1} atomic absorption lines, and CO overtone absorption bands.  Br$\gamma$ is seen weakly in emission.  Spectral template fitting suggests a best-fit spectral type of M2$\pm$2, with A$_V\sim$12$\pm$2 magnitudes, and r$_K\sim$0.4$\pm$0.2.   The derived A$_V$=12 supports the interpretation above of the continuum change at 1.7~$\mu$m.  As an example of the best-fit spectrum and the associated $\chi^2$ uncertainty from the spectral template fitting procedure,  we show the results for CN~2 in Figs.~\ref{bestfit} and \ref{chisq}, respectively.  For brevity, similar plots for other stars analyzed with spectral template fitting will not be explicitly presented.   The $\chi^2$ surfaces in Fig.~\ref{chisq} are defined by a minimization search performed on A$_V$ and r$_K$ and, for these sources, assumes a flat temperature distribution rather than one defined by a single T$_{dust}$ black-body.\footnotemark\footnotetext{It was found that the model fits were insensitive to the T$_{dust}$ value (for the relatively small r$_K$ values encountered) and that a flat distribution gave slightly improved fits.}  Here we show the $\chi^2$ = 1.01, 2.0, and 3.0 surfaces defining the uncertainties on A$_V$ and r$_K$ for several stellar spectral types.  The 1$\sigma$ deviation approximately corresponds to the  $\chi^2\sim$3.0 surface.

From its association with large-scale bipolar HH jets, its Br$\gamma$ emission, and its K-band thermal excess, we confirm the source as an active, young stellar object (YSO).  The strength of its water vapor absorption with respect to its relatively weak CO absorption are somewhat inconsistent suggesting that it possesses a circumstellar disk with a cool extended atmosphere.  Such an inconsistency has previously been seen in other T~Tauri stars with the above interpretation (Shiba et al. 1993).  The fact that the two HH flows are not co-linear suggests that it may be a close binary.  CN~2 should be designated an accreting CTTS star and warrants further investigation.

\subsection{CN~3N}
This source is the third compact nebula found in the region by Movsessian et al. (2003).  There are in fact two stars associated with the nebulosity (CN~3N and 3S, separated by $\sim$27$''$) with the northern source appearing to be both an IRAS source (IRAS20585+5222) and an MSX source (G091.8730+04.2583).  The fan-shaped nebula seen by Movsessian et al. (2003) appears more diffuse in our images (Fig.~\ref{ids2}) and could be equally well associated with either star.  In the NIR, CN~3N is brighter than CN~3S, the opposite is true in the optical.  

CN~3N is reddened with J-K$\sim$3.3.  In the JHKcc diagram (Fig.~\ref{ccdiag}), it lies on the reddening vector from the location of early-type ZAMS stars and its location is therefore consistent with it being either a reddened A--F main sequence dwarf or a reddened CTTS star.  

The spectrum of CN~3N shows a rising red continuum with few features.  The change in continuum slope at ~$\sim$1.7~$\mu$m is indicative of overlying visual extinction.  Present in the spectrum are weak Br(10-4), Br$\gamma$, \ion{Na}{1}, \ion{Ca}{1}, \ion{Mg}{1}, and CO absorption, the latter being more apparent in the expanded K band plot in Fig.~\ref{sp-1-k}.  These features rule out the possibility of CN~3N being a reddened early-type star.  The spectral model fitting suggest that CN~3N is a G8 star with an A$_V$=20$\pm$3, and r$_K$=0.2$\pm$0.1.  An inconsistency, however, is that the weakness of continuum change at 1.7~$\mu$m suggests a smaller A$_V$ value.  

CN~3N is an embedded YSO (dereddening to the CTTS locus in Fig.~\ref{ccdiag} gives A$_V\sim$20) with a small K-band thermal excess.  It is associated with a weak IRAS source, 20585+5222, with the only good quality (q=3) detection at 25$\mu$m (0.46~Jy).   Unusually, CN~3N shows no Br$\gamma$ emission rather weak absorption.  The weak absorption spectrum can be interpreted as from a reddened G-type dwarf although this is somewhat inconsistent with the association with an IRAS point-source unless a weakly emitting dusty circumstellar disk is present.

\subsection{CN~3S}
CN~3S is a moderately reddened star with J-K$\sim$2.3 and in the JHKcc diagram (Fig.~\ref{ccdiag}), its location is consistent with it being either a reddened, late-type main sequence dwarf (A$_V\sim$8), a more heavily reddened G--F dwarf (A$_V\sim$11), or a reddened CTTS stars (A$_V\sim$8).  

CN~3S shows water vapor absorption (indicated by the dip in continuum level longward of 2.3~$\mu$m) together with \ion{Na}{1}, \ion{Ca}{1} and CO absorption features.  There is no compelling evidence for Br$\gamma$ in either emission or absorption.  The result of the spectral template fitting suggests that CN~3S is of spectral type M0 with an A$_V$=15$\pm$3, and r$_K$=0.1$\pm$0.1.  The dramatic change in continuum slope at 1.7~$\mu$m supports such a level of visual extinction.  The above atomic and molecular features suggests that CN3~S is not of a G--F dwarf classification.

CN~3S appears somewhat similar in spectral characteristics to CN~3N.  However, no K-band thermal excess and no IRAS association argues for a reddened dwarf classification.  If CN~3S is a reddened M dwarf (we assume an M0~V), then it should have an M$_K\sim$5.2 and hence a (reddening corrected) distance modulus of 4.5.  This corresponds to $\sim$80pc and is not reconcilable with its reddened nature.  It has an m$_K$=10.82 and to be behind the L1003 molecular cloud and reddened by it, a distance of at least 800pc.  For this to occur, CN~3S would need to be an early-A dwarf with an M$_K$=1.3.   This is inconsistent with the spectral features observed.  Alternatively, if we place the source at 800pc, and assume an M$_K$=5.2, CN~3S would be around 5 magnitudes over-luminous for its spectral type.   In this case, CN~3S would undoubtably be a young star contracting along its Hayashi track.  It is likely however, that such a star would still be actively accreting resulting in Br$\gamma$ emission.  Also, it would have significant circumstellar dust resulting in a thermal excess and perhaps IRAS detection.   Perhaps optical spectroscopy and longer wavelength NIR photometry will resolve this issue.

\subsection{CN~6}
The original CN~6 designation refers to a faint nebulous object seen in our optical Subaru images (Fig.~\ref{ids3}).  In the NIR however, it is the southern of two relatively bright (m$_K\sim$11.3 and 8.7), reddened (J-K$\sim$3.9 and 5.9) stars which we henceforth refer to as CN~6 and CN~6N, respectively.  These objects are separated by $\sim$30$''$ but only CN~6 appears nebulous at K with a morphology suggestive of a monopolar cavity with its axis pointing to the south-west.  CN~6 possesses a significant thermal excess and is consistent with a reddened (A$_V\sim$11) CTTS at the extreme of the MCH97 CTTS locus.

CN~6 has a declining red continuum with perhaps weak water vapor absorption bands. Br$\gamma$ absorption (see Fig.~\ref{sp-1-k}), and CO overtone absorption are probably present.  It is not clear if \ion{Na}{1} and \ion{Ca}{1} are present but if so, they are very weak.  Due to the weakness of all these features, spectral template fitting was not attempted.

It is clear that CN~6 is a YSO.  Its morphology suggests the presence (at some time) of a wide-angle molecular outflow and its large K-band excess and (faint) optical appearance implies the presence of a circumstellar disk.  The weakness of the Ca and Na lines,  no detectable Mg line, and weak CO bandhead absorption, suggests it is perhaps of spectral type mid-G to mid-K.  We adopt the classification of an embedded CTTS for CN~6.

\subsection{CN~6N}
As we have seen, CN~6N is a highly reddened (J-K$\sim$5.9) star. It lies slightly to the left of the upper reddening vector in Fig.~\ref{ccdiag} although within the uncertainties, it can be considered on the reddening vector.  Dereddening by A$_V\sim$28 puts it on the giant branch at round early-M type.   It shows no K-band thermal excess emission.  

The CO overtone bandheads in CN~6N are very deep and both \ion{Na}{1} and \ion{Ca}{1} are seen in absorption (see Fig.~\ref{sp-1}).  CN~6N has a red rising continuum and there is no evidence for Br$\gamma$ in either absorption or emission.  Water vapor absorption bands are present and consistent with it being a late-type background giant.  The only other possibility is that it is FUor-like since these are the NIR characteristic of FUors (Reipurth \& Aspin 1997).  Spectral template fitting results in a spectral type of early-M~III, an A$_V$=30$\pm$3, and an r$_K$=0.1$\pm$0.1.  Assuming CN~6N is an M3~III giant with an M$_K$=--4 and A$_V$=30 (A$_K$=3), this would place it at a distance of 850pc, consistent with its reddened nature.   Any later than M3~III would place CN~6N foreground to the L1003 molecular cloud.   No thermal excess argues against it being a FUor-like object in L1003.

\subsection{CN~7}
CN7 lies close to the tail of the Braid Nebula and in the same general area as CN~2, CN~8, and IRAS~20588+5215 (see Fig.~\ref{ids2}).  In the optical, CN~7 itself appears to possess a tail of nebulosity which is seen curving to the north-west.  It is also associated with a small compact nebulous source about 4$''$ to the south-east.  At K, CN~7 is a bright (m$_K\sim$10.8) point-source.  Its NIR colors suggest that it is somewhat reddened (J-K$\sim$1.5) and in the JHKcc diagram it lies close to both the lower reddening vector and the MCH97 CTTS locus.  Dereddening to the ZAMS implies an A$_V\sim$8 and an early-F classification.  Dereddening to the MCH97 CTTS locus would result in a considerable smaller value of A$_V\sim1$.

The H-band spectrum of CN~7 shows rising then declining continuum effected by extinction at the longward end.  The spectrum in the K-band continues to decline.  In terms of spectral features, CN~7 has \ion{Mg}{1} and \ion{Al}{1} absorption in the H-band and \ion{Na}{1}, \ion{Ca}{1} and CO overtone absorption in the K-band.  Br$\gamma$ is seen in neither absorption or emission.  A spectral type between around M1 is suggested by the spectral template fitting with an A$_V$=6$\pm$2 and r$_K$=0.3$\pm$0.1.

CN~7 is certainly a YSO.  The association of CN~7 with the curving cavity-wall like nebulosity suggests that it has had a molecular outflow at some time in the past.   Its spectral features resemble those of an early-M dwarf with several magnitudes of visual extinction and some K-band veiling.  No Br$\gamma$ emission suggests it is not actively accreting.  

\subsection{CN~8}
CN~8 exhibits relatively bright curving optical nebulosity extending some 15$''$ away from the star to the north-east and, to a lesser extent, to the north-west.   It is a bright point-source in the NIR with an m$_K\sim$9.8 and a J-K$\sim$1.9 and, as we mentioned above, is within 90$''$ of three other young nebulous objects (CN~2, CN~7, and IRAS~20588+5215, see Fig.~\ref{ids2}).  CN~8 is not however included in the IRAS point-source catalog.  In the JHKcc diagram (Fig.~\ref{ccdiag}), CN~8 possesses a small thermal excess lying close to the extreme of the MCH97 CTTS locus.  

The most prominent feature of the H and K spectra of CN~8 is Br$\gamma$ emission together with higher order H lines emission (i.e. Br10, Br11, etc).  There is perhaps a hint of water vapor absorption and few other features are evident.  Very weak \ion{Na}{1} and \ion{Ca}{1} absorption may be present (see Fig.~\ref{sp-1-k}).  The continuum flux rises monotonically to the red.   The weakness of the spectral features present means we could not apply our spectral template fitting to CN~8.
   
Clearly CN~8 is an accreting CTTS.  This interpretation is supported by its Br$\gamma$ emission (implying accretion activity), the K-band thermal excess (implying the presence of heated circumstellar dust), and the cavity-like nebulosity (implying the presence, at some time, of a wide-angle molecular outflow).
 
\subsection{Cyg~19}
Cyg~19 is a bright optical and NIR source with associated curving nebulosity in both wavelength regimes (Fig.~\ref{ids3}).  It was included in the H$\alpha$ emission line catalog of Melikian et al. (1996) from which it gets its identification.  In both the optical and NIR, the bright point-source (m$_K\sim$10.0) has several fainter objects within a few arcseconds and it may be a member of a multiple system.  The star is only moderately reddened (J-K$\sim$1.4) and it is located between the reddening vectors in Fig.~\ref{ccdiag}, close to the MCH97 T~Tauri star locus.  If we deredden its location onto the ZAMS we find it corresponds to either an early-G dwarf with an A$_V\sim$7, or an M3--4 dwarf with A$_V\sim$3.  

The continuum flux in the spectrum of Cyg~19 first increases in strength and then declines to the edge of the K-band indicative of overlying extinction.  Upon the continuum are several other features, specifically, atomic \ion{Mg}{1} (three lines), \ion{Al}{1} and \ion{Si}{1}, \ion{Na}{1}, and \ion{Ca}{1} absorption, and molecular CO overtone band absorption.  No H emission or absorption features are seen.  Spectral template fitting suggest these features are best represented by a star of spectral type K7, an A$_V$=5$\pm$2 and an r$_K$=0.1$\pm$0.05.

The curving nebulosity around Cyg~19 is reminiscent of an outflow cavity seen close to pole-on and observed via scattered light.   This is further supported by the optical visibility of the source.  This association is enough to support the conclusion that Cyg~19 is a sporadically accreting YSO.   However, the addition of the H$\alpha$ detection by Melikian et al. (1996) is added confirmation.  We do not observe H emission lines, however, which may imply that the accretion process varies significantly with time.  

\subsection{HH381~IRS}
This source is located approximately 19$'$ west of the Braid star and was first studied in the NIR by Reipurth \& Aspin (1997). They found that HH381~IRS had a 2~$\mu$m spectrum similar in characteristics to the FUor class of eruptive variables.  Our imaging shows it to be associated with a bipolar conical nebula in both the optical and NIR  (Fig.~\ref{ids3}) which has been recently shown to have brightened significantly between image taken in 1989 and 1990 (Magakian et al. 2007\footnotemark\footnotetext{available at http://www.iaus243.org/IMG/pdf/Magakian1.pdf}).  HH381~IRS is named after its association with the HH objects HH381 and 382 although it is not transparent that this young star is truly the source of these shock excited gaseous knots. Finally, HH381~IRS has an IRAS identification, 20568+5217, with a large 60$\mu$m flux (10~Jy).

HH381~IRS is bright at K (m$_K\sim$8) and is significantly reddened (J-K$\sim$3.2).  In Fig.~\ref{ccdiag}, its location suggests it has a significant thermal excess and it is located to the right of the reddening vector from the extreme of the MCH97 CTTS locus.  

The HK spectrum of HH381~IRS is dominated by probably a combination of water vapor absorption bands and extinction effects, and deep CO overtone absorption is present.  The K-band spectrum is very similar to that found by Reipurth \& Aspin (1997) and shows few atomic features apart from weak Br$\gamma$, \ion{Na}{1}, \ion{Ca}{1} and \ion{Mg}{1} absorption lines.  It was not possible to obtain a spectral type by template fitting due to the weakness of these lines.

HH381~IRS is definitely a YSO similar in several ways to the FUor variables.  The strong IRAS detection implies a significant dust mass around the source likely in the form of a circumstellar disk since bipolar conical cavities are visible in the optical and NIR.  It is strange that we see do not see Br$\gamma$ emission yet this is also the case in the FUors V1057~Cyg, V1515~Cyg, and V1735~Cyg and it likely due to the disk dominating the emission and absorption characteristics in the NIR.  We classify HH381~IRS as an embedded FUor similar to the Braid star.

\subsection{HH627-STAR}
This bright optical star is located some 90$''$ south-east of CN~6 and seems to be  associated with HH627 located 20$''$ south-east since the bow shock morphology of HH627 point approximately away from this star.  It is due to this association that we included the star in our NIR spectroscopic survey and henceforth refer to it as HH627-STAR.

HH627~STAR is relatively unreddened with a J-K$\sim$0.4 (see Fig.~\ref{ids2}). It lies close to the unreddened ZAMS in the JHKcc diagram (Fig.~\ref{ccdiag}) at the location of a late-G to early-K dwarf.   Its NIR spectrum shows a monotonically declining continuum with weak absorption lines of \ion{H}{1}, \ion{Mg}{1}, \ion{Na}{1}, \ion{Ca}{1}, plus CO overtone band absorption.  The depths of CO and Br$\gamma$ absorption constrain the spectral type to G6 to K0.

There is nothing to suggest that this star is a YSO other than its intimate location near HH627.   In fact, all the evidence points to it being a foreground object.  Its NIR colors and spectral features both suggest it is a late-G to early-K foreground dwarf meaning the association with HH627 is purely fortuitous.  If we assume HH627-STAR is a G8~V then for m$_K$=10.36, an absolute K-band magnitude M$_K$=3.9, and A$_V$=0, we can place HH627-STAR at a distance of around 200~pc.

\subsection{IRAS~20591+5214}
This object is one of two IRAS sources in the region found to be associated with optical and/or NIR sources.  IRAS~20591+5214, herein referred to as IRAS~14, was not detected in our optical images but is seen as a compact diffuse nebula with point-source (m$_K\sim$14) in the NIR (Fig.~\ref{ids3}).  The IRAS detection are labelled 'L' as upper limits at all but 25~$\mu$m where the flux is 0.4~Jy.  A peculiarity associated with this source is that SIMBAD\footnotemark \footnotetext{This research has made use of the SIMBAD database, operated at CDS, Strasbourg, France.} shows IRAS~14 to have a V magnitude of 14.3 and a designation of Cyg~21 from the Melikian et al. (1996) catalog of H$\alpha$ emission line stars.  This is unusual since our deep Subaru optical images show no source present.  Since IRAS~14 was not detected in our J band WFCAM images, we cannot place it on the color-color diagram.  However, with an H-K=1.4 it is at least reddened to the extent of A$_V\sim$15--20 magnitudes.  

IRAS~14's NIR spectrum is noisy due to the fainter nature of the source, however, Br$\gamma$ emission is present.  The emission peak just longward of the $v=2-0$ CO overtone bandhead at 2.294~$\mu$m is not intrinsic to the source but an artifact of poor telluric correction.  

We can conclude that IRAS~14 is an accreting YSO undergoing accretion activity and must possess a dusty circumstellar envelope/disk producing far-IR flux.   It is probable that the source is highly variable in nature.  The fact that it was once optically visible together with the lack of a strong IRAS detection suggests that it perhaps possesses  variable circumstellar extinction.

\subsection{IRAS~20588+5215N}
IRAS~20588+5215 (herein referred to as IRAS~15) is bright in both the optical and NIR.  It is located near CN~2, 7, and 8 and upon closer inspection we see that it is a close double source with a separation of $\sim$5$''$.  Both components (herein designated IRAS~15N and 15S) appear of similar brightness in the optical but in the NIR, IRAS~15N is significantly brighter.  

2MASS resolved the two stars and found that IRAS~15N had J-K=1.75.  In our JHKcc diagram (Fig.~\ref{ccdiag}) it appears close to the MCH97 CTTS locus and has a definite K-band thermal excess.  It was saturated in our recent WFCAM images and therefore we cannot comment on its variability.  

The HK spectrum of IRAS~15N shows several interesting features.  Br$\gamma$ and the lower transition of H are strongly in emission as are the CO overtone bandheads.  The atomic \ion{Na}{1}, \ion{Ca}{1}, and \ion{Mg}{1} lines are however, in absorption.  The turnover in continuum slope at 1.7~$\mu$m suggests a reddened nature.  The emission characteristics of the HK spectrum of IRAS~15N makes spectral template fitting impossible.

IRAS~15N is a highly active accreting accreting CTTS with a significant K-band thermal excess.  It must also possess a hot, dense inner circumstellar disk which produces the CO bandhead emission.  

\subsection{IRAS~20588+5215S}
The 2MASS photometry of IRAS~15S gives J-K=0.61.  In our JHKcc diagram (Fig.~\ref{ccdiag}) IRAS~15S is located on the unreddened ZAMS at the location of an early-K dwarf or mid-G giant suggesting perhaps that IRAS~15S is an unreddened foreground object. 

The HK spectrum of IRAS~15S shows a turnover at 1.7~$\mu$m with a declining continuum to both blue and red wavelengths suggestive of significant overlying extinction.  No statistically significant atomic absorption lines are present.  One interesting fact is that the molecular hydrogen, H$_2$, v=1-0 S(1) line at 2.122~$\mu$m is seen weakly in emission. 

From the above photometry, IRAS~15S could be interpreted as a foreground object with a derived spectral type of early-K~V or mid-G~III.  If this is correct, it is difficult to explain the H$_2$ emission in the spectrum. 

\section{Summary}
Of the 16 sources observed spectroscopically, three are either foreground or background luminosity class V dwarfs.  One additional object is a background luminosity class III giant.   Thus, 12 sources are YSOs.  The properties of all stars observed are summarized in Table~\ref{whatn}.  What we find is:

\begin{itemize}
\item The YSOs investigated here, have characteristics showing them to be either highly active accreting CTTSs or relatively inactive, embedded, young low-mass stars.  Two FUor-like objects are also present. 

\item Of the 12 YSOs, six were detected by IRAS and eight exhibit K-band thermal excess emission.

\item All YSOs observed are associated with optical and/or NIR reflection nebulosity.

\item Four YSOs possess HH flows/jets.

\item Six YSOs are actively accreting (as indicated by Br$\gamma$ line emission). 

\item Of the four YSOs with HH flows/jets, all have a K-band thermal excess emission.  Two of these YSOs have an emission spectrum while the other two only possess molecular absorption bands.  

\item Two YSOs are FUor-like in their spectral characteristics with both being associated with HH objects.  However, only one of these objects is an IRAS source. 

\item The derived visual extinction towards all source varies from an A$_V$ of 0 to a maximum of 30 magnitudes.  

\item The K-band veiling of the YSO stellar spectra is in the range r$_K$=0.1 to 0.4.  These are relatively modest values and is perhaps related to the maximum veiling present that allows spectral template fitting with this spectral resolution.  Another way of expressing this, is that the K-band spectral features fitted in the spectral modeling are rendered too weak to be fitted in spectra with r$_K>$0.4.

\item Spectral template fitting has resulted in estimates of spectral type, visual extinction (A$_V$), and K-band veiling (r$_K$) for four YSOs and show a range in these parameters of spectral type G8--M2, A$_V$ of 5--20 magnitudes, r$_K$ of 0.1--0.4.  The results were not sensitive to the T$_{dust}$ adopted (due to the small r$_K$ derived) with a flat temperature distribution being found to produce the bes
t-fit to the observations.   This flat distribution suggest that the dust emission is not well represented by a single temperature but by a range of values.

\end{itemize}

We conclude by noting that the Braid Nebula region of the L~1003 molecular cloud in the Cyg~OB7 association is a highly active star forming region with what is likely to be a significant population of YSOs.  Active molecular outflows are implied by nebula morphology, HH flows are seen in shock excited optical and NIR emission, and a range of YSO characteristics are observed.   We note that the region bares some resemblance to the NGC~1333 star forming region studied in the optical and NIR by Aspin, Sandell, \& Russell (1994), Bally, Devine, \& Reipurth (1996), and Lada, Alves, \& Lada (1996) although the Braid region perhaps contains a smaller population of very young embedded protostellar sources.   However, as in NGC1333, the Braid Nebula star formation region exhibits multiple molecular outflow sources, a plethora of HH objects, and a variety of young star masses and ages.

\vspace{0.3cm}

{\bf Acknowledgments} 

The United Kingdom Infrared Telescope is operated by the Joint Astronomy Centre on behalf of the Science and Technology Facilities Council of the U.K.  We thank A.~Nord, L.~Rizzi, \& T.~Carroll for assistance in obtain these observations.   We also thank the University of Hawaii Time Allocation Committee for allocating the nights during which these observations were made.  The authors wish to recognize and acknowledge the very significant cultural role and reverence that the summit of Mauna Kea has always had within the indigenous Hawaiian community. We are most fortunate to have the opportunity to conduct observations from this sacred mountain.  This publication makes use of data products from the Two Micron All Sky Survey, which is a joint project of the University of Massachusetts and the Infrared Processing and Analysis Center/California Institute of Technology, funded by the National Aeronautics and Space Administration (NASA) and the National Science Foundation.  CA acknowledges page charges support from a small grant allocation made by the American Astronomical Society.

% REFERENCES 

\clearpage 
%############################################################################################

% Table 1
\begin{deluxetable}{llllll}
\tabletypesize{\scriptsize}
\tablecaption{Object Coordinates and Identification\label{sourceci}}
\tablewidth{0pc}
%\rotate
\tablehead{
\colhead{Object} & 
\colhead{2MASS} & 
\colhead{IRAS} &
\colhead{MSX} &
\colhead{R.A.} &
\colhead{Decl.} \\
\colhead{Designation} &
\colhead{ID} &
\colhead{ID} &
\colhead{ID} &
\colhead{(J2000)} &
\colhead{(J2000)}
}

\startdata
Braid~star & --               & --         & --                & 21:00:25.38 & +52:30:15.5 \\
CN~1       & 21003517+5233244 & 20590+5221 & --                & 21:00:35.17 & +52:33:24.4 \\
CN~1S      & 21003552+5233043 & --         & --                & 21:00:35.52 & +52:33:04.3 \\
CN~2       & 21001656+5226230 & --         & --                & 21:00:16.56 & +52:26:23.0 \\
CN~3N      & 21000376+5234290 & 20585+5222 &  091.8730+04.2583 & 21:00:03.76 & +52:34:29.0 \\
CN~3S      & 21000508+5234049 & --         & --                & 21:00:05.08 & +52:34:04.9 \\
CN~6       & 20594071+5234135 & --         & --                & 20:59:40.71 & +52:34:13.5 \\
CN~6N      & 20594000+5234361 & --         & --                & 20:59:40.00 & +52:34:36.1 \\
CN~7       & 21001725+5228253 & --         & --                & 21:00:17.25 & +52:28:25.3 \\
CN~8       & 21001903+5227281 & --         & --                & 21:00:19.03 & +52:27:28.1 \\
Cyg~19     & 21001159+5218172 & --         & --                & 21:00:11.59 & +52:18:17.2 \\
HH381~IRS  & 20582109+5229277 & 20568+5217 & --                & 20:58:21.09 & +52:29:27.7 \\
HH627-STAR & 20594446+5233213 & --         & --                & 20:59:44.46 & +52:33:21.3 \\
IRAS~14\tablenotemark{a}    & 21004237+5226007 & 20591+5214 & --                & 21:00:42.37 & +52:26:0.07 \\
%IRAS~15N\&S\tablenotemark{a} & 21002140+5227094 & 20588+5215 & 21:00:21.40 &
%+52:27:09.4 \\
IRAS~15N\tablenotemark{b}   & 21002140+5227094 & 20588+5215 &  --  & 21:00:21.40 & +52:27:09.4 \\
IRAS~15S\tablenotemark{b}   & 21002113+5227052 & 20588+5215 &  --  & 21:00:21.13 & +52:27:05.2 \\
\enddata
\tablenotetext{a}{IRAS~14 is IRAS~20591+5214.}
\tablenotetext{b}{IRAS~15N and 15S are the unresolved IRAS source 20588+5215.}
\end{deluxetable}
\clearpage

%############################################################################################

% Table 2
\begin{deluxetable}{lllllll}
\tabletypesize{\scriptsize}
\tablecaption{Near-IR 2MASS and WFCAM Photometry\label{sourceph}}
\tablewidth{0pc}
\rotate
\tablehead{
\colhead{Object} & 
\colhead{m$_J$\tablenotemark{a}} &
\colhead{m$_H$\tablenotemark{a}} &
\colhead{m$_K$\tablenotemark{a}} & 
\colhead{J-H\tablenotemark{a}} &
\colhead{H-K\tablenotemark{a}} &
\colhead{J-K\tablenotemark{a}} \\
\colhead{ID} &
\colhead{(mags)} &
\colhead{(mags)} &
\colhead{(mags)} &
\colhead{(mags)} &
\colhead{(mags)} & 
\colhead{(mags)}
}

\startdata
Braid~star & --(16.46$\pm$0.03) & --(15.34$\pm$0.02) & --($>$11.5\tablenotemark{b}) & 1.12$\pm$0.04 & 4.89$\pm$0.5 & 6.01$\pm$0.5 \\
CN~1 & 15.27$\pm$0.06(15.07$\pm$0.02) & 13.86$\pm$0.04(13.90$\pm$0.02) & 12.52$\pm$0.03(12.60$\pm$0.02) & 1.41$\pm$0.08(1.17$\pm$0.03) & 1.34$\pm$0.05(1.30$\pm$0.03) & 2.75$\pm$0.07(2.47$\pm$0.03) \\
CN~1S & 11.56$\pm$0.02(sat\tablenotemark{c}) & 10.73$\pm$0.02(sat)  & 10.28$\pm$0.02(sat)  & 0.83$\pm$0.03 & 0.45$\pm$0.03 & 1.28$\pm$0.03 \\
CN~2 & 12.59$\pm$0.02(12.73$\pm$0.02) & 11.47$\pm$0.02(sat) & 10.67$\pm$0.02(10.86$\pm$0.02) & 1.12$\pm$0.03 & 0.80$\pm$0.03 & 1.92$\pm$0.03 \\ 
CN~3N & 13.32$\pm$0.03(12.77) & 11.26$\pm$0.02(sat) & 9.98$\pm$0.02(sat) & 2.06$\pm$0.04 & 1.28$\pm$0.03 & 3.34$\pm$0.04 \\
CN~3S & 13.16$\pm$0.02(12.95$\pm$0.02) & 11.61$\pm$0.02(sat) & 10.82$\pm$0.02(sat) & 1.55$\pm$0.03 & 0.79$\pm$0.03 & 2.34$\pm$0.03 \\
CN~6  & 15.21$\pm$0.06(14.93$\pm$0.02) & 13.03$\pm$0.03(12.88$\pm$0.02) & 11.32$\pm$0.03(11.02$\pm$0.02) & 2.18$\pm$0.07(2.05$\pm$0.03) &
1.71$\pm$0.05(1.86$\pm$0.03) & 3.89$\pm$0.08(3.91$\pm$0.03) \\
CN~6N & 14.57$\pm$0.03(14.53$\pm$0.02) & 10.61$\pm$0.02(sat) & 8.66$\pm$0.02(sat) & 3.96$\pm$0.04 & 1.95$\pm$0.03 & 5.91$\pm$0.04 \\
CN~7 & 12.28$\pm$0.02(sat) & 11.35$\pm$0.02(sat) & 10.82$\pm$0.02(sat) & 0.93$\pm$0.03 & 0.53$\pm$0.03 & 1.46$\pm$0.03 \\
CN~8 & 11.69$\pm$0.02(sat) & 10.63$\pm$0.02(sat) & 9.81$\pm$0.02(sat) & 1.06$\pm$0.03 & 0.82$\pm$0.03 & 1.88$\pm$0.03 \\
Cyg~19 & 11.38$\pm$0.02(sat) & 10.48$\pm$0.02(sat) & 10.02$\pm$0.02(sat) & 0.90$\pm$0.03 & 0.46$\pm$0.03 & 1.36$\pm$0.03 \\
HH381~IRS & 11.54$\pm$0.03(sat) & 9.81$\pm$0.03(sat) & 8.3$\pm$0.02(sat) & 1.73$\pm$0.05 & 1.51$\pm$0.04 & 3.24$\pm$0.05 \\
HH627-STAR & 10.77$\pm$0.02(sat) & 10.42$\pm$0.02(sat) & 10.36$\pm$0.02(sat) & 0.35$\pm$0.03 & 0.06$\pm$0.03 & 0.41$\pm$0.03 \\
IRAS~14 & --(--) & 15.50$\pm$0.08(17.95$\pm$0.04) & 14.1$\pm$0.06(14.72$\pm$0.03) & --(--) & 1.40$\pm$0.12 & -- \\
%IRAS~15N\&S\tablenotemark{e,f} & 11.15$\pm$0.03(sat) &10.22$\pm$0.03(sat) & 9.4$\pm$0.04(sat) & 0.93$\pm$0.05 & 0.82$\pm$0.06 & 1.75$\pm$0.06 \\
IRAS~15N & 11.15$\pm$0.03(sat) & 10.22$\pm$0.03(sat) & 9.40$\pm$0.04(sat) & 0.93$\pm$0.05 & 0.82$\pm$0.05 & 1.75$\pm$0.05 \\
IRAS~15S & 12.45$\pm$0.03(12.40$\pm$0.02)  & 11.96$\pm$0.04(12.10$\pm$0.02) & 11.84$\pm$0.03(11.92$\pm$0.02) & 0.49$\pm$0.05(0.30$\pm$0.03) & 0.12$\pm$0.05(0.18$\pm$0.03) & 0.61$\pm$0.05(0.48$\pm$0.05) \\
\enddata
\tablenotetext{a}{2MASS photometry(WFCAM photometry). Errors are taken from
2MASS and WFCAM catalogs.}
\tablenotetext{b}{K band photometry an upper limit from WFCAM saturation magnitude.}
\tablenotetext{c}{(sat) means saturated in WFCAM images.}
\end{deluxetable}
\clearpage

%############################################################################################

% Table 3
\begin{deluxetable}{lcrrrrcrrrrrr}
\tabletypesize{\scriptsize}
\tablecaption{Spectral Features and Equivalent Widths\label{whatp}}
\tablewidth{0pc}
\rotate
\tablehead{
\colhead{Object} & 
\colhead{S:N$_H$\tablenotemark{a}} &
\colhead{\ion{Mg}{1}} & 
\colhead{\ion{Si}{1}\tablenotemark{b}} & 
\colhead{\ion{Al}{1}\tablenotemark{b}} & 
\colhead{\ion{Mg}{1}} & 
\colhead{S:N$_K$\tablenotemark{a}} &
\colhead{H$_2$} & 
\colhead{Br$\gamma$} &
\colhead{\ion{Na}{1}} &
\colhead{\ion{Ca}{1}} &
\colhead{\ion{Mg}{1}} & 
\colhead{$v=2-0$ CO\tablenotemark{c}} \\
\colhead{Name} &
\colhead{} &
\colhead{1.505\tablenotemark{d}} &
\colhead{1.668} &
\colhead{1.674} &
\colhead{1.711} &
\colhead{} &
\colhead{2.122} &
\colhead{2.166} &
\colhead{2.208} &
\colhead{2.262} &
\colhead{2.281} &
\colhead{2.294} \\
\colhead{} &
\colhead{} &
\colhead{(\AA)\tablenotemark{e}} &
\colhead{(\AA)} &
\colhead{(\AA)} &
\colhead{(\AA)} &
\colhead{} &
\colhead{(\AA)} &
\colhead{(\AA)} &
\colhead{(\AA)} &
\colhead{(\AA)} &
\colhead{(\AA)} &
\colhead{(\AA)}}

\startdata
%           S:NH   MgI   SiI    AlI    MgI   S:NK  H2      Brg     NaI    CaI    MgI      CO      
%                 1504  1668   1674   1711        2122    2166    2208   2262   2281    2294    
Braid~star & 7   & --  & --   & --   & --  & 30  & --6   & --    & ?    & ?    & ?\tablenotemark{f} & 18   \\
CN~1       & 11  & --  & --   & --   & --  & 30  & --    & --3   & ?    & 9    & ?     & 2   \\
CN~1S      & 60  & --  & --   & --   & --  & 100 & --    & 18    & --   & --   & --    & --  \\
CN~2       & 55  & 5   & 3    & 2    & --  & 75  & --    & --1   & 2    & 3    & 1     & 4   \\
CN~3N      & 120 & 4   & ?    & ?    & 1   & 190 & --    & 1     & 1    & 1    & 1     & 2   \\
CN~3S      & 80  & 3   & 2    & 2    & 2   & 100 & --    & --    & 2    & 3    & 1     & 3   \\
CN~6       & 25  & --  & --   & --   & --  & 25  & --    & 7     & 8    & 5    & --    & 9   \\
CN~6N      & 50  & 4   & 3    & 1    & 3   & 140 & --    & --    & 3    & 3    & 1     & 17  \\
CN~7       & 75  & 3   & 2    & 2    & 3   & 75  & --    & --    & 2    & 1    & --    & 3   \\
CN~8       & 11  & --  & --   & --   & --  & 25  & --    & --4   & 2    & 8    & 1     & 4   \\
Cyg~19     & 70  & 4   & 1    & 2    & 3   & 120 & --    & --    & 3    & 3    & 1     & 4   \\
HH381~IRS  & 80  & 1   & --   & --   & --  & 120 & --    & --    & 1    & 1    & 1     & 7   \\
HH627-STAR & 95  & 3   & ?    & ?    & 1   & 90  & --    & 3     & 2    & 2    & 1     & 3   \\
IRAS~14    & 10  & --  & --   & --   & --  & 10  & --    & --    & --   & --   & --    & --  \\
IRAS~15N   & 50  & --  & --   & --   & 1   & 50  & --    & --6   & 1    & 2    & 1     & --6 \\
IRAS~15S   & 40  & 4   & 2    & --   & 2   & 30  & --7   & --    & --   & --   & --    & --  \\
\enddata
\tablenotetext{a}{Signal to noise in the reduced spectrum in the H or K band estimated from the r,s noise in a featureless region.}
\tablenotetext{b}{\ion{Si}{1} and \ion{Al}{1} are blended at the spectral resolution of UIST. We have estimated their relative contribution by considering the shape of the line profile.}
\tablenotetext{c}{The v=2-0~CO bandhead W$_{\lambda}$ is determined between 2.289 and 2.306~$\mu$m}
\tablenotetext{d}{Wavelength of spectral feature in microns.}
\tablenotetext{e}{Typical uncertainty on W$_{\lambda}$ values is $\pm$0.5~\AA.}
\tablenotetext{f}{? means possible present but W$_{\lambda}$ $<$1~\AA.}
\end{deluxetable}
\clearpage

%############################################################################################

% Table 4
\begin{deluxetable}{llllllllllllll}
\tabletypesize{\scriptsize}
\tablecaption{Nature of Observed Sources\label{whatn}}
\tablewidth{0pc}
\rotate
\tablehead{
\colhead{Object} &
\colhead{Optical} & 
\colhead{Near-IR} &
\colhead{IRAS} &
\colhead{Thermal\tablenotemark{a}} &
\colhead{Assoc.\tablenotemark{b}} &
\colhead{Emission\tablenotemark{c}} &
\colhead{HH\tablenotemark{d}} &
\colhead{Young\tablenotemark{e}} & 
\colhead{Evol.\tablenotemark{f}} &
\colhead{Spect.\tablenotemark{g}} &
\colhead{A$_V$\tablenotemark{h}} & 
\colhead{A$_V$\tablenotemark{i}} & 
\colhead{r$_K$\tablenotemark{j}} \\
\colhead{Name} &
\colhead{Source?} &
\colhead{Source?} &
\colhead{Source?} &
\colhead{Excess?} &
\colhead{Neb.?} &
\colhead{Spect?} &
\colhead{jet/flow} &
\colhead{Star?} &
\colhead{Class} & 
\colhead{Type} &
\colhead{(mags)} &
\colhead{(mags)} &
\colhead{}}

\startdata
%            Opt   NIR   IRAS   TE    REF   EM    HH    YSO    Cla     SpT     Av1   Av2     rK
Braid~star &  no & yes &  no & yes & yes &  no & yes & yes  &  YSO  & --    & --  & --    & --  \\
CN~1       & yes & yes & yes & yes & yes & yes & yes & yes  &  YSO  & --    & --  & --    & --  \\
CN~1S      & yes & yes &  no &  no &  no &  no & no  &  no  & dwarf & B8V   & --  & 6     & --  \\
CN~2       & yes & yes &  no & yes & yes & yes & yes & yes  &  YSO  & M2    & 12  & 12    & 0.4 \\
CN~3N      & yes & yes & yes &  no & yes &  no & no  & yes  &  YSO  & G8    & 20  & 20    & 0.2 \\
CN~3S      & yes & yes &  no &  no & ?\tablenotemark{k}   &  no & no  & ?    &   ?   & M0    & 15  & 11    & 0.1 \\
CN~6       & yes & yes &  no & yes & yes &  no & no  & yes  &  YSO  & --    & --  & 11    & --  \\
CN~6N      &  no & yes &  no &  no &  no &  no & no  & no   & giant & M3III & 30  & 28    & 0.1 \\
CN~7       & yes & yes &  no &  no & yes &  no & no  & yes  &  YSO  & M1    & 6   & 8     & 0.3 \\
CN~8       & yes & yes &  no & yes & yes & yes & no  & yes  &  YSO  & --    & --  & --    & --  \\
Cyg~19     & yes & yes &  no &  no & yes &  no & no  & yes  &  YSO  & K7    & 5   & 7     & 0.1 \\
HH381~IRS  & yes & yes & yes & yes & yes &  no & yes & yes  &  YSO & --    & --  & --    & --  \\
HH627-STAR & yes & yes &  no &  no &  no &  no & no  &  no  & dwarf & G8V   & --  & 0     & --  \\
IRAS~14    &  no & yes & yes & X   & yes &  no & no  & yes  &  YSO  & --    & --  & $<$18 & --  \\
IRAS~15N   & yes & yes & yes & yes & yes & yes & no  & yes  &  YSO  & --    & --  & --    & --  \\
IRAS~15S   & yes & yes & yes & yes & yes & yes & no  & yes  &  YSO  & --    & --  & 0     & --  \\
\enddata

\tablenotetext{a}{Near-IR K-band thermal excess from JHKcc diagram (Fig.~\ref{ccdiag})}
\tablenotetext{b}{Source has associated reflection nebulosity.}
\tablenotetext{c}{Emission lines in H and/or K band spectra.}
\tablenotetext{d}{Driving with an Herbig-Haro flow or jet.}
\tablenotetext{e}{Implied from thermal excess, association with reflection nebulosity or emission spectrum.}
\tablenotetext{f}{Spectral type derived from spectral template fitting.}
\tablenotetext{g}{Spectral classification, see Lada (1997).}
\tablenotetext{h}{A$_V$ estimated from spectral template fitting.}
\tablenotetext{i}{A$_V$ estimates from dereddening to ZAMS in Fig.~\ref{ccdiag}.}
\tablenotetext{j}{K-band veiling derived from spectral template fitting.}
\tablenotetext{k}{A ? indicates unresolved classification or ambiguous association.}
\end{deluxetable}
\clearpage

%############################################################################################
% Figure area
\clearpage
%############################################################################################
% Figure IDs1
\begin{figure*}[tb]
\epsscale{1.0}
\plotone{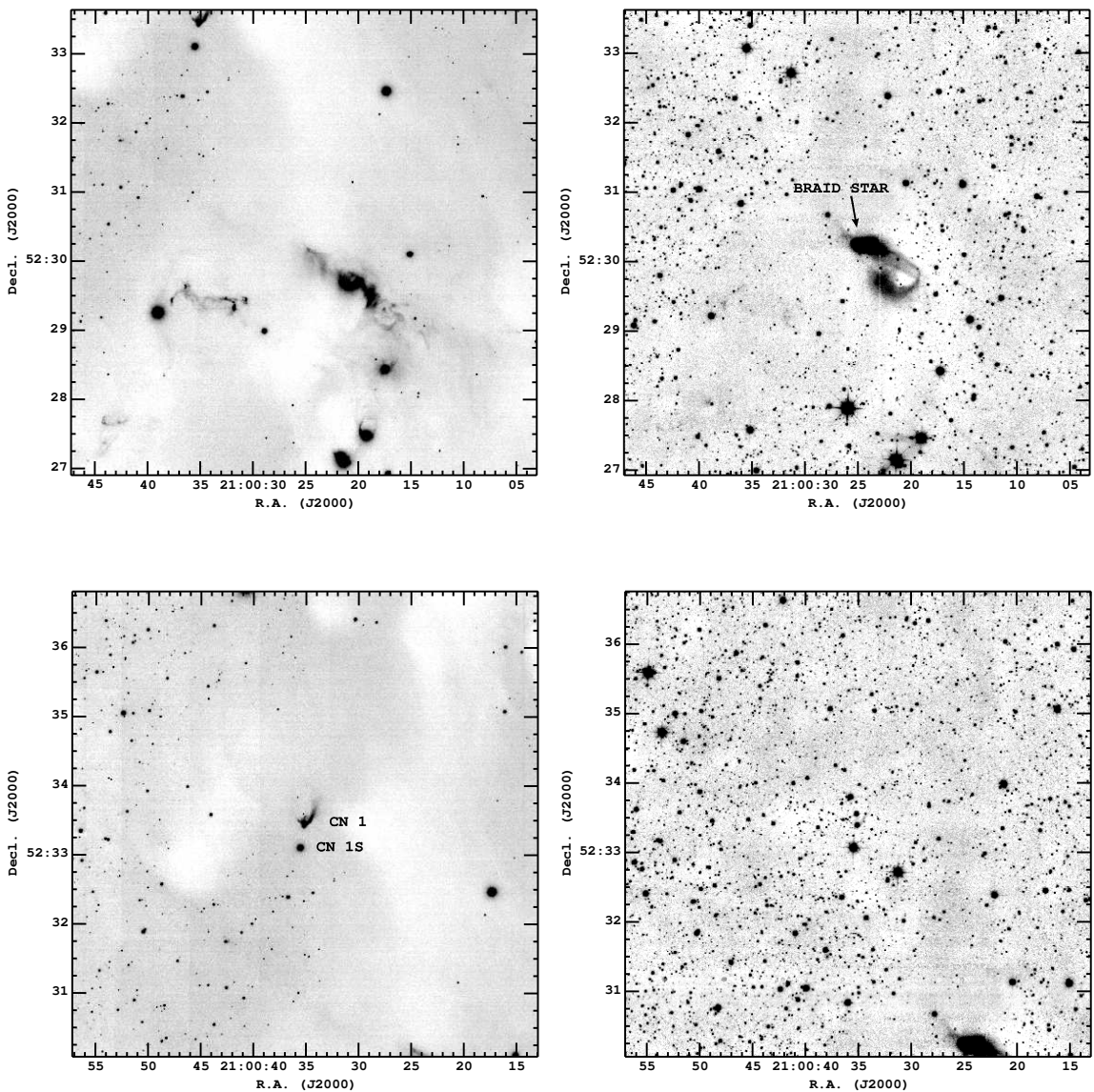} 
\caption{Optical R$_C$ (left) and NIR K band (right) images of the fields containing the target stars in Cygnus~OB~7.   The stars for which we have obtained NIR spectroscopy are identified by name.  In these images, north is at the top and east to the left.  The field of view of each image is $\sim$7.5$'$ square.  The R$_C$ image is from the Subaru 8-meter telescope using Suprimecam while the K band image is from the UKIRT 4-meter telescope using WFCAM.
\label{ids1}}
\end{figure*}
\clearpage
%############################################################################################
% Figure IDs2
\begin{figure*}[tb]
\epsscale{1.0}
\plotone{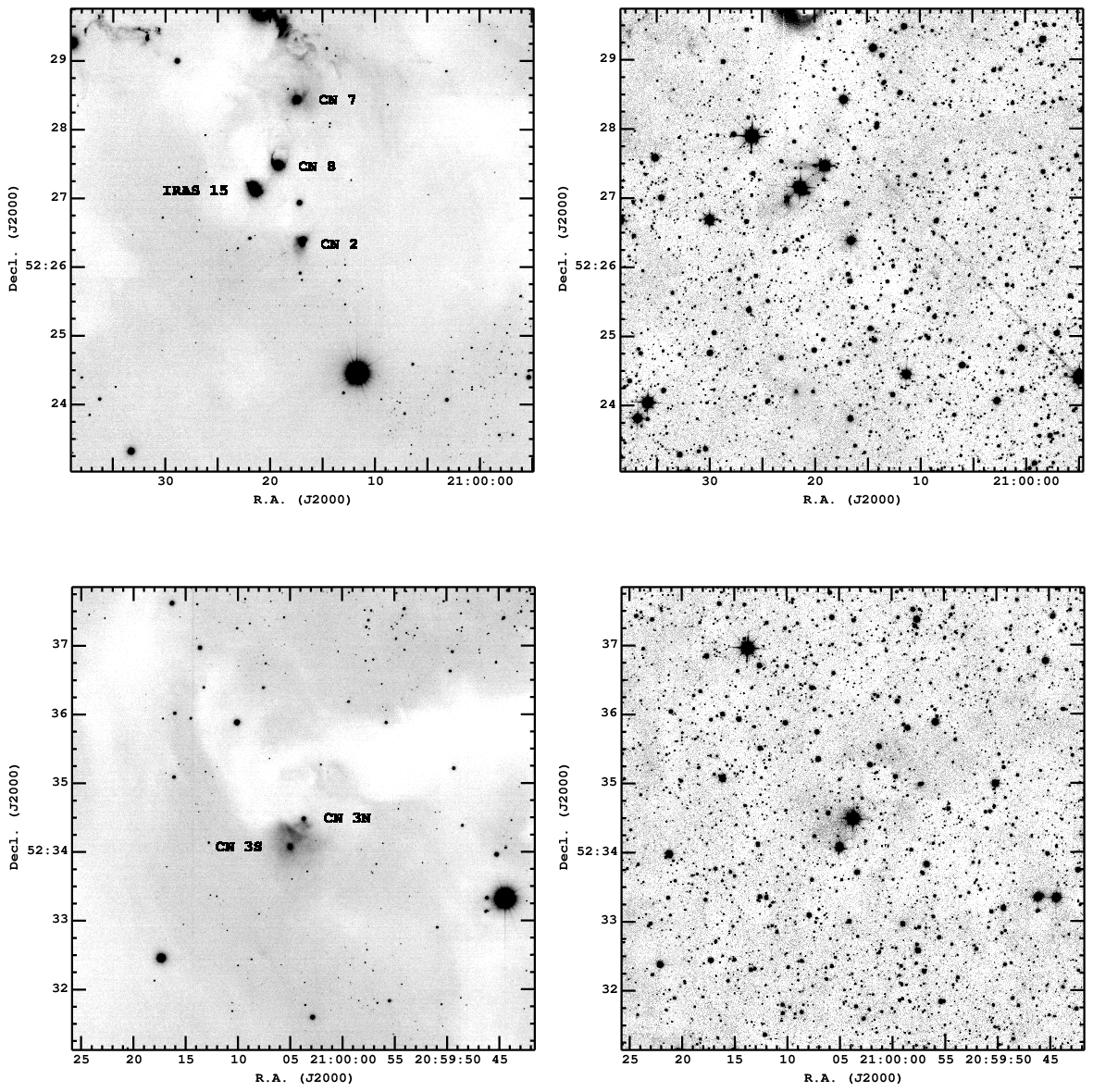} 
\addtocounter{figure}{-1}
\caption{continued...
\label{ids2}}
\end{figure*}
\clearpage
%############################################################################################
% Figure IDs3
\begin{figure*}[tb] 
\epsscale{1.0}
\plotone{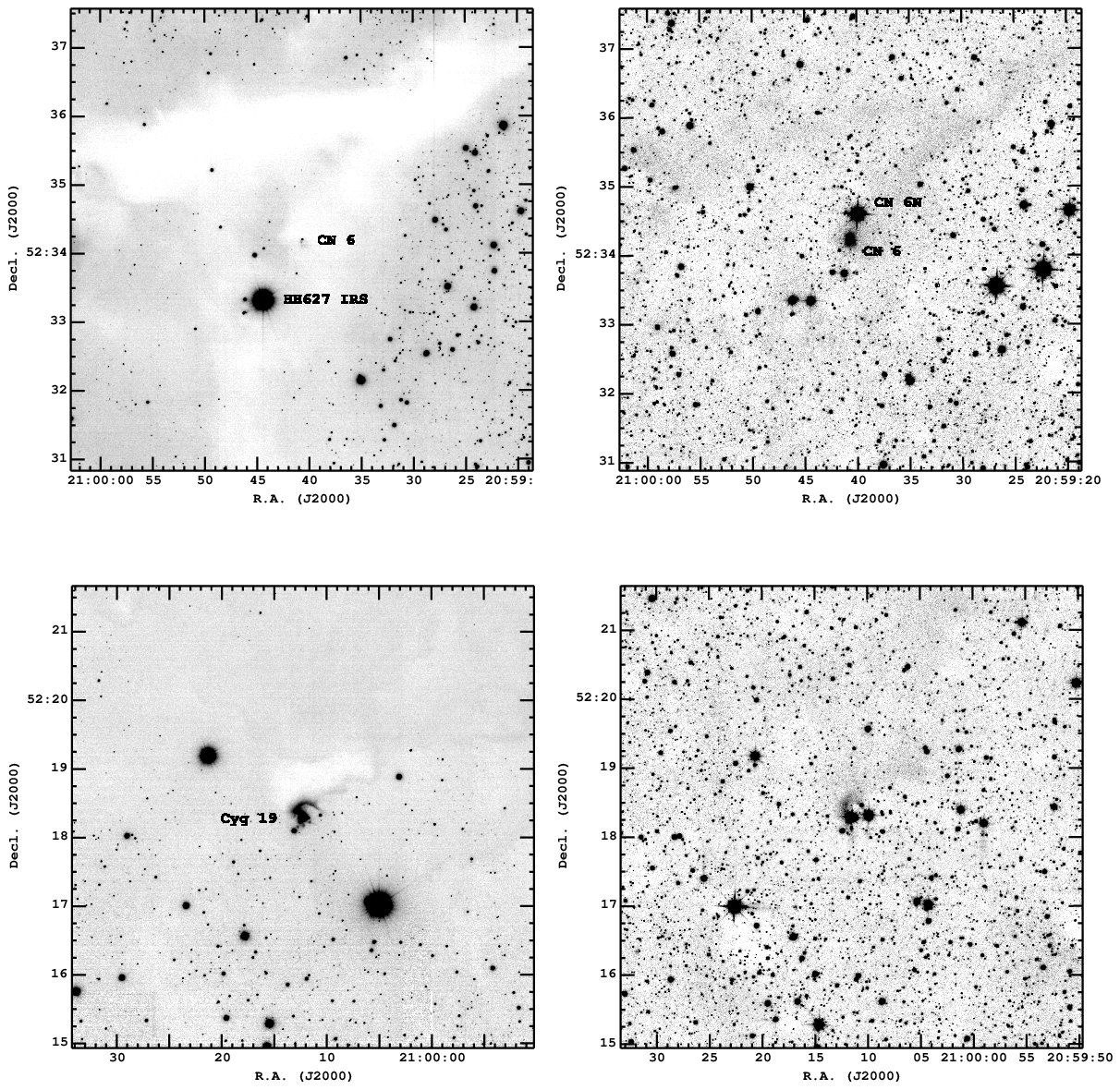} 
\addtocounter{figure}{-1}
\caption{continued...
\label{ids3}}
\end{figure*}
\clearpage
%############################################################################################
% Figure IDs4
\begin{figure*}[tb] 
\epsscale{1.0}
\plotone{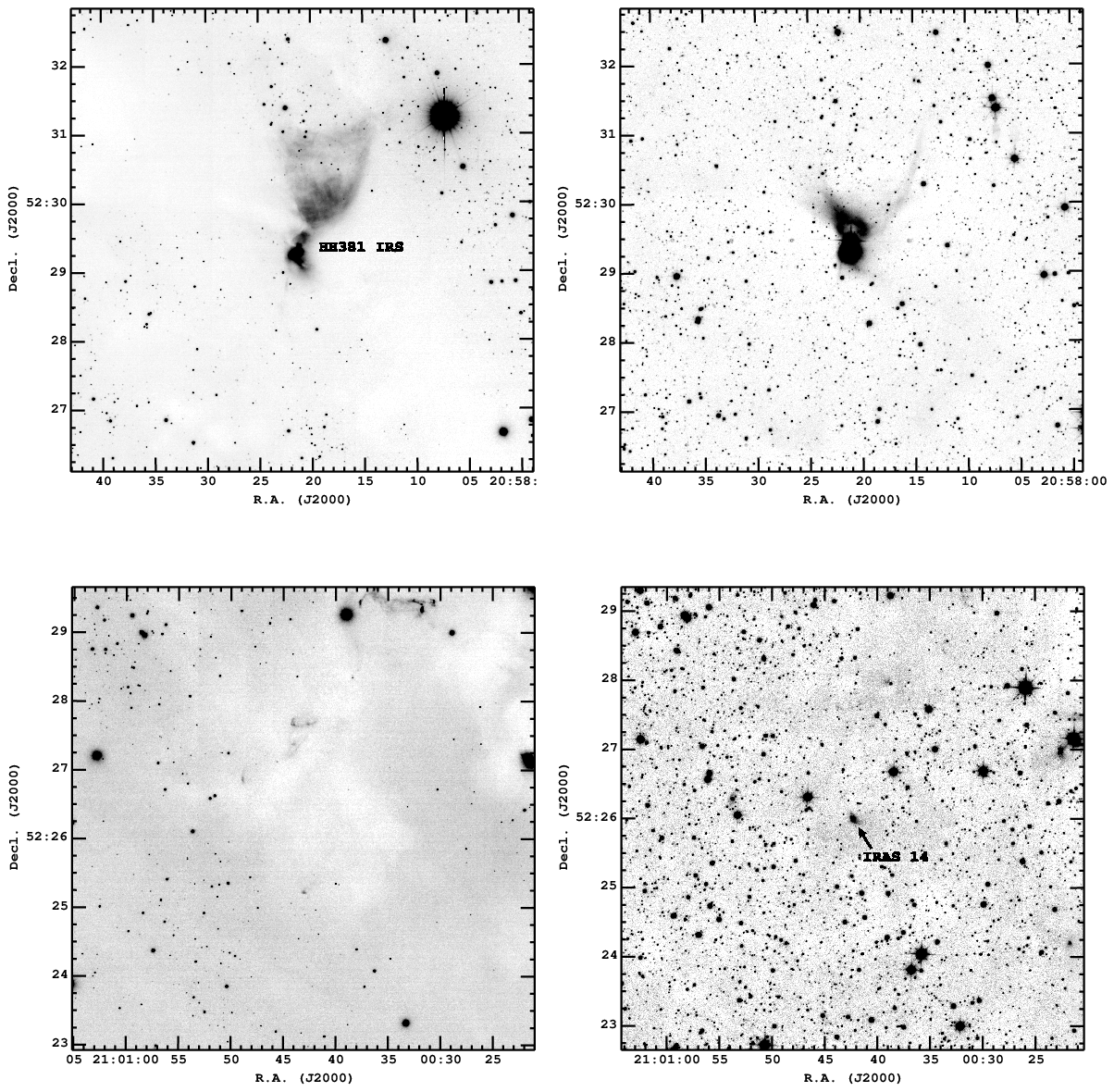} 
\addtocounter{figure}{-1}
\caption{continued...
\label{ids4}}
\end{figure*}
\clearpage

%############################################################################################
% Figure cc
\begin{figure*}[tb] 
\epsscale{1.0}
\plotone{braid-nirspectra-2mass.ps} 
\caption{J-H vs H-K color--color diagram for the sources observed.  The photometry is given in Table~\ref{sourceph} and is taken from 2MASS.  The solid lines with open and solid points are the loci of main sequence dwarfs and giant stars colors.  The two straight solid lines are reddening vectors plotted from the extremes of the dwarf/giant loci and are A$_V$=30 magnitudes.  The dashed line is the locus of classical T~Tauri stars from Meyer, Calvet, \& Hillenbrand (1997) and the dot-dashed line is a reddening vector (for A$_V$=20) from the extreme of this locus.  The object observed spectroscopically are shown as large filled circles with associated labels.  Representative error bars for both J-H and H-K are shown at J-H=0.5, H-K=2.0.  
\label{ccdiag}}
\end{figure*}
\clearpage

%############################################################################################
% Figure sp-1
\begin{figure*}[tb] 
\epsscale{1.0}
\plotone{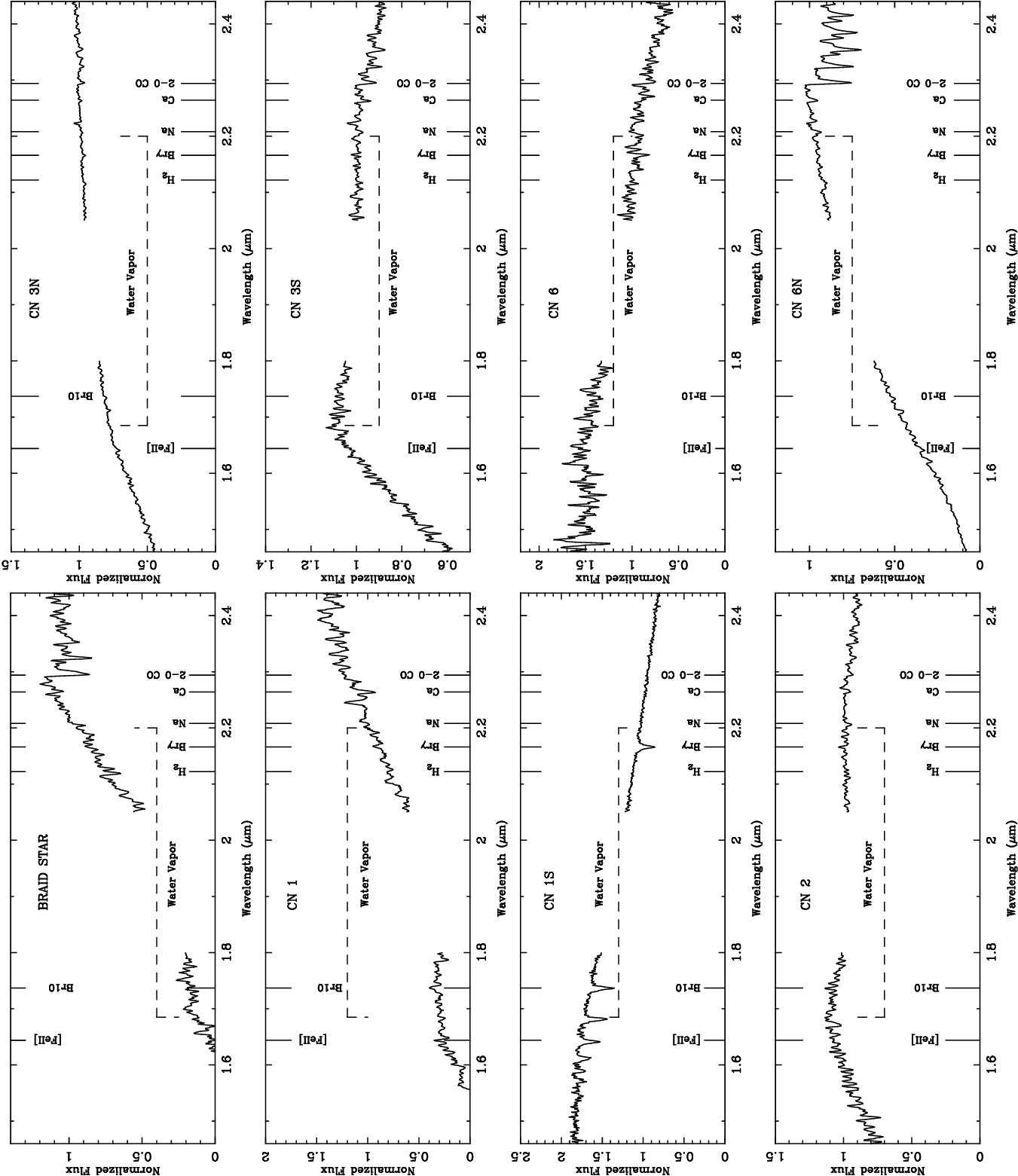} 
\caption{H and K band spectra of the sources observed.  The plots have been scaled to show the broad shape of the continuum flux rather than faint line emission/absorption.  The important atomic lines and molecular bands are identified.  The spectra have been scaled to unity at 2.2~$\mu$m.  
\label{sp-1}}
\end{figure*}
\clearpage

%############################################################################################
% Figure sp-2
\begin{figure*}[tb] 
\epsscale{1.0}
\plotone{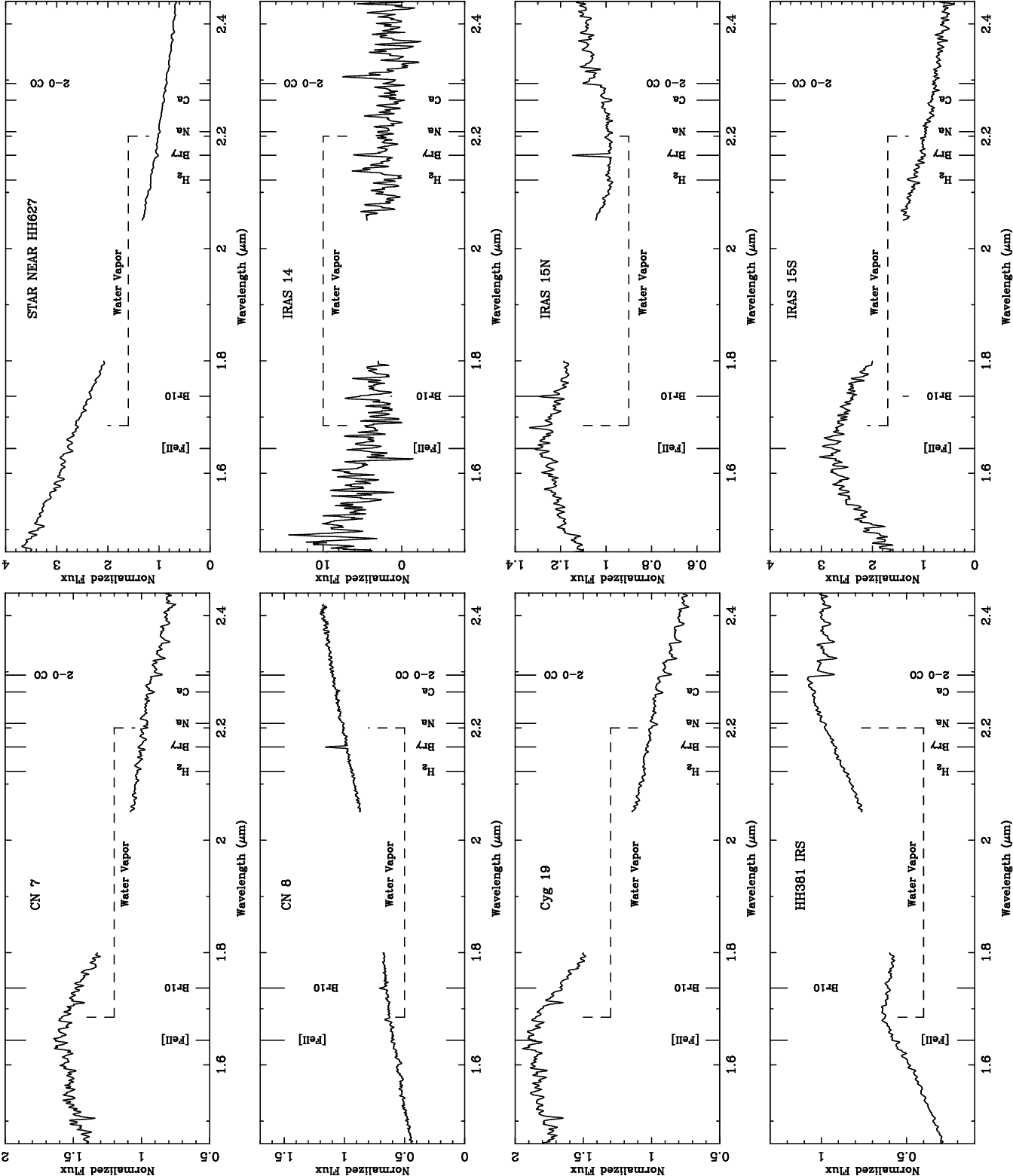} 
\addtocounter{figure}{-1}
\caption{continued...
\label{sp-2}}
\end{figure*}
\clearpage

%############################################################################################
% Figure sp-1-h
\begin{figure*}[tb] 
\epsscale{1.0}
\plotone{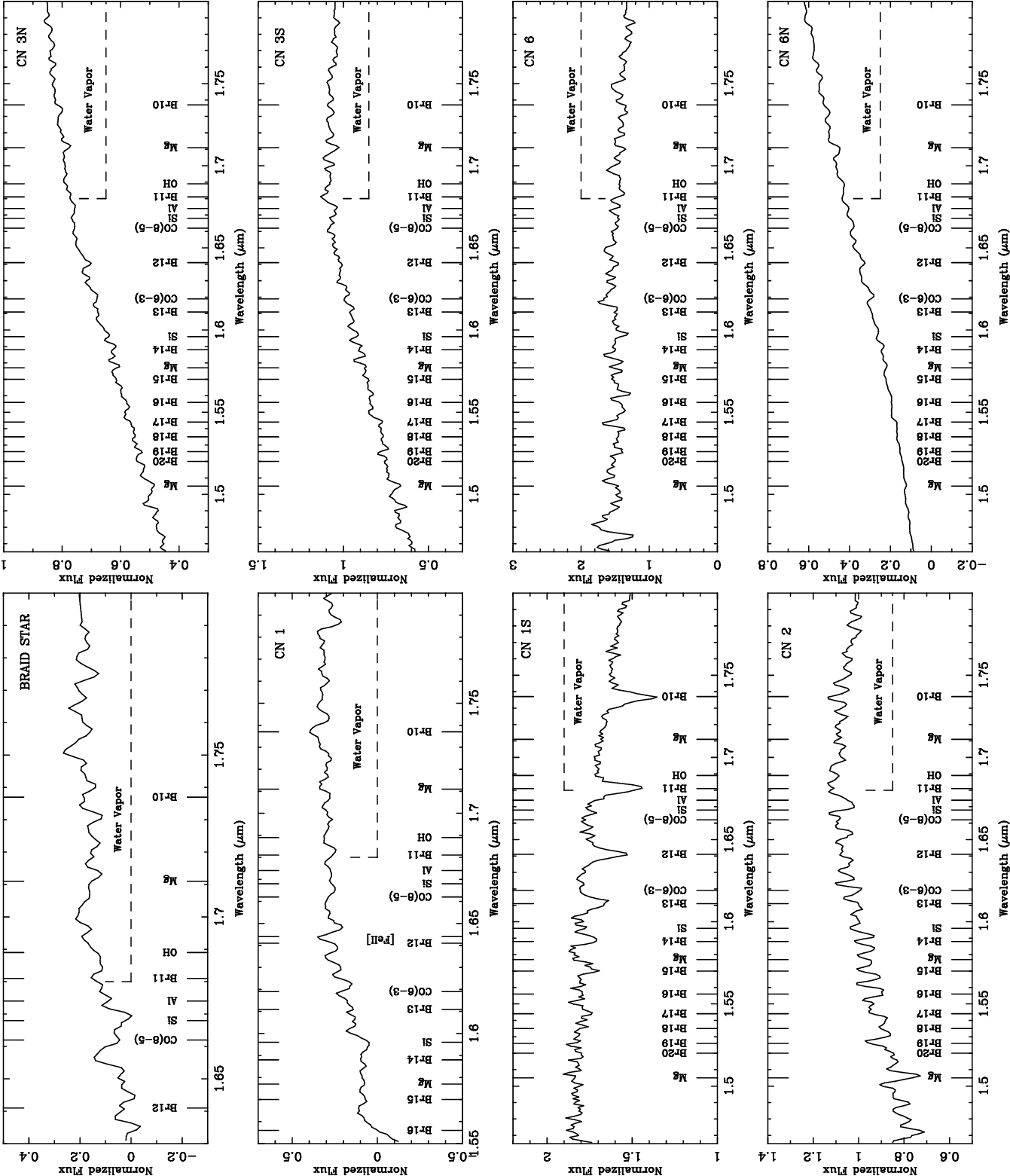} 
%\addtocounter{figure}{-1}
\caption{An expanded view of the H band region of the spectra.  All line/bands in the passband are identified.  The wavelength range of some plot is truncated at the short-wavelength end of the spectrum at the wavelength where the observed flux goes to zero.  
\label{sp-1-h}}
\end{figure*}
\clearpage

%############################################################################################
% Figure sp-2-h
\begin{figure*}[tb] 
\epsscale{1.0}
\plotone{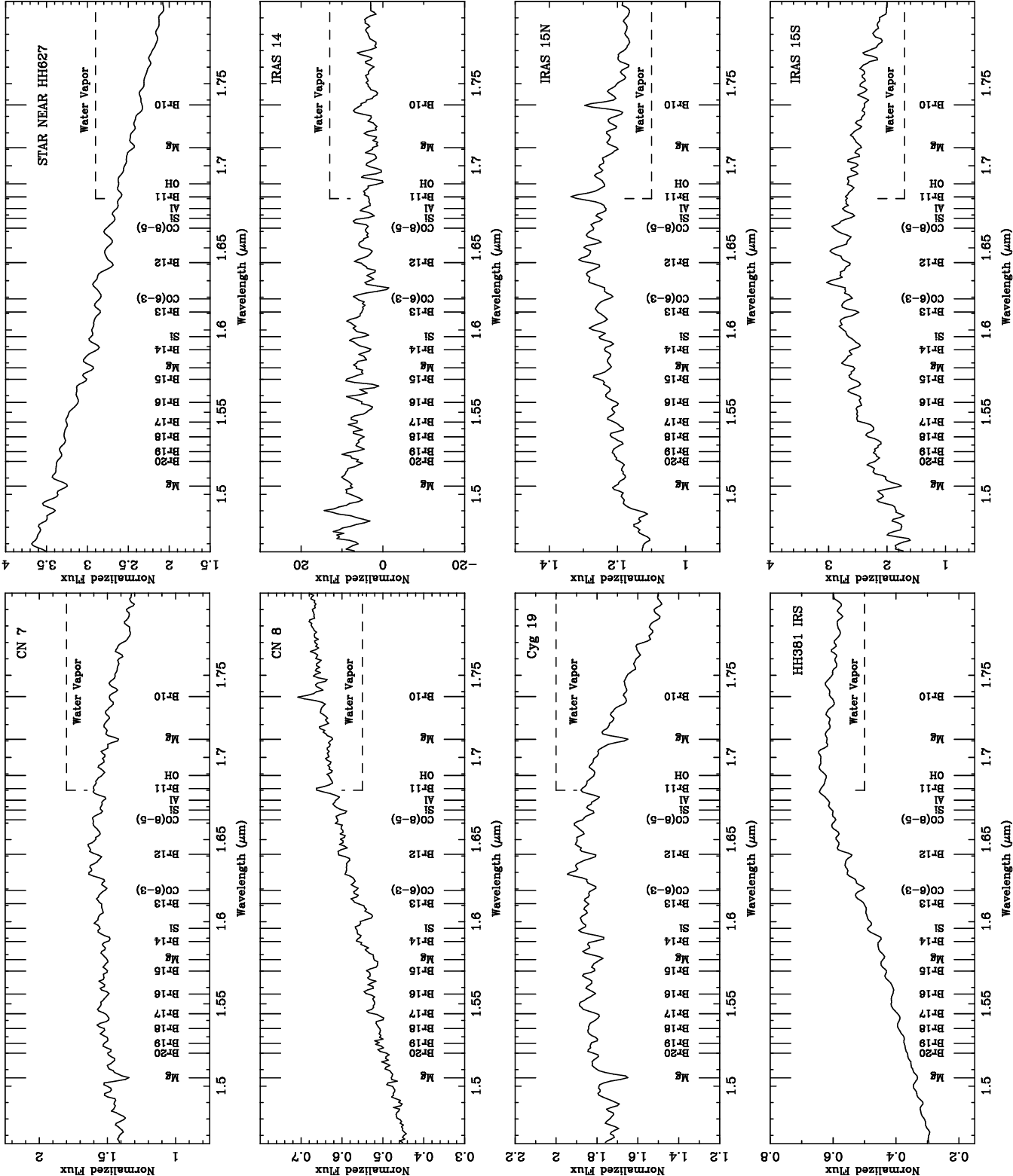} 
\addtocounter{figure}{-1}
\caption{continued...
\label{sp-2-h}}
\end{figure*}
\clearpage

%############################################################################################
% Figure sp-1-k
\begin{figure*}[tb] 
\epsscale{1.0}
\plotone{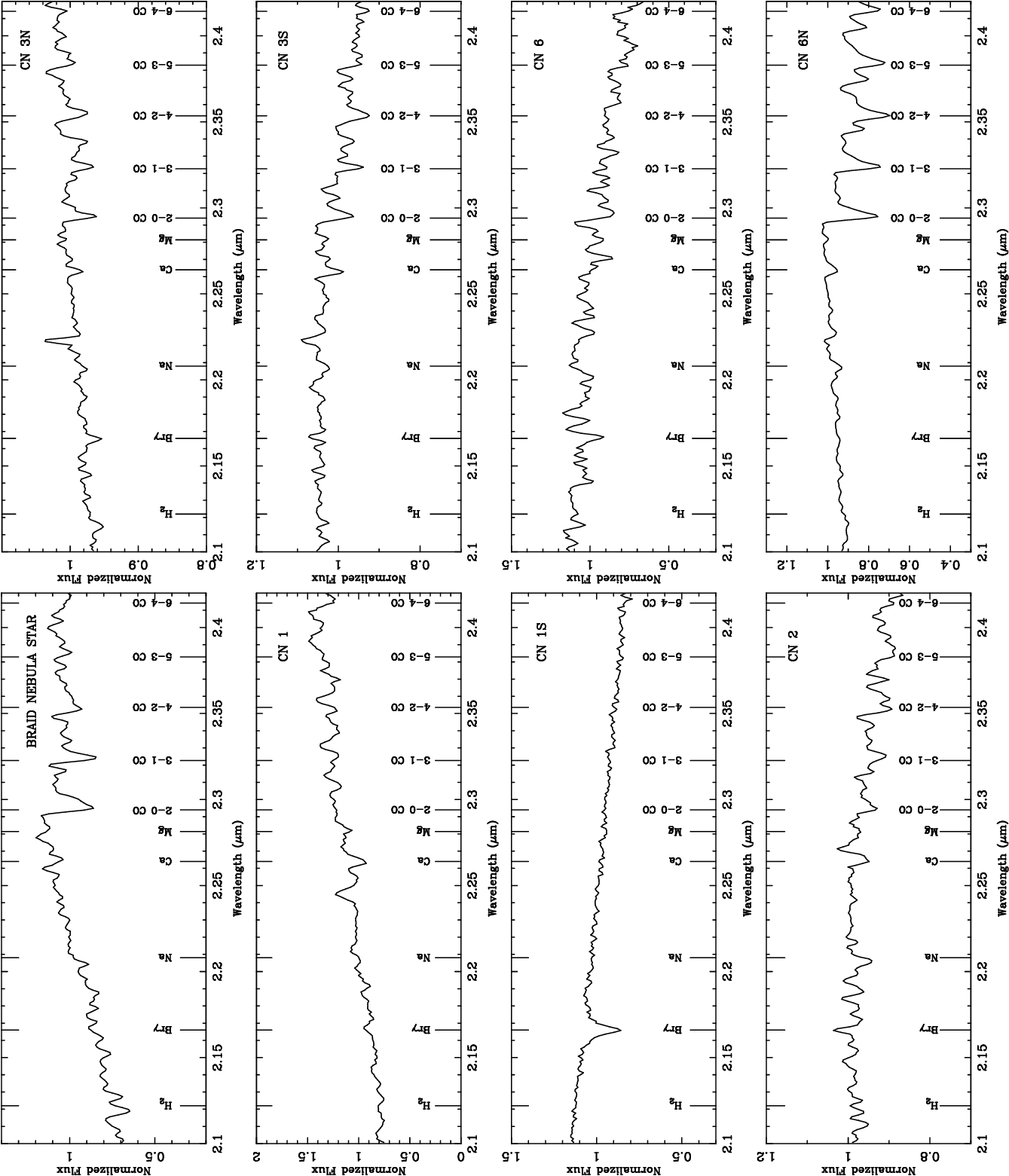} 
%\addtocounter{figure}{-1}
\caption{An expanded view of the K band region of the spectra.  All line/bands in the passband are identified.  
\label{sp-1-k}}
\end{figure*}
\clearpage

%############################################################################################
% Figure sp-2-k
\begin{figure*}[tb] 
\epsscale{1.0}
\plotone{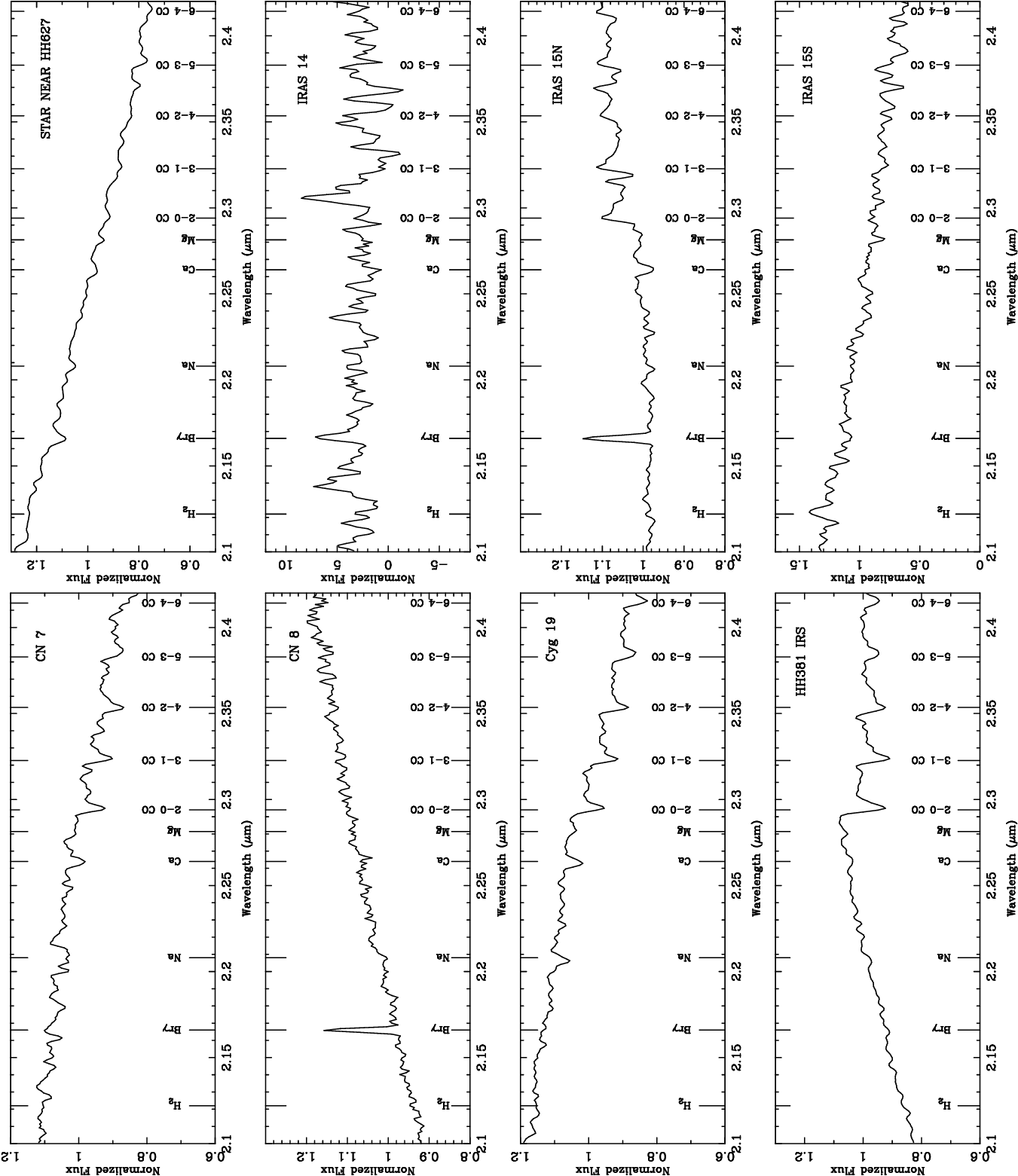} 
\addtocounter{figure}{-1}
\caption{continued...
\label{sp-2-k}}
\end{figure*}
\clearpage

%############################################################################################
% Figure best-fit CN2
\clearpage

%############################################################################################
% Figure chi-squared CN2
\clearpage

\end{document}